\documentclass{aa}
\usepackage{graphicx}
\usepackage{txfonts}
\usepackage[dvipsnames]{xcolor}
\usepackage{float}
\begin{document} 

   \title{A new method for age-dating the formation of bars in disc galaxies}

   \subtitle{The TIMER view on NGC1433's old bar and the inside-out growth of its nuclear disc}

%   \author{C. de Sá-Freitas
%           \inst{1}
%           D. Gadotti\inst{1}\fnmsep\thanks{Just to show the usage
%           of the elements in the author field}
%           }

   \author{Camila de Sá-Freitas
          \inst{1}\thanks{\email{camila.desafreitas@eso.org}}
          \and
          Francesca Fragkoudi \inst{1,2,3}
          \and
          Dimitri A. Gadotti\inst{1,4}
          \and
          Jesús Falcón-Barroso\inst{5,6}
          \and
          Adrian Bittner\inst{1,7}
          \and
          Patricia Sánchez-Blázquez\inst{8, 9}
          \and
          Glenn van de Ven\inst{10}
          \and
          Rebekka Bieri\inst{3}
          \and
          Lodovico Coccato\inst{1}
          \and
          Paula Coelho\inst{11}
          \and
          Katja Fahrion\inst{12}
          \and
          Geraldo Gonçalves\inst{11}
          \and
          Taehyun Kim\inst{13}
          \and
          Adriana de Lorenzo-Cáceres\inst{5,6}
          \and
          Marie Martig\inst{14}
          \and
          Ignacio Martín-Navarro\inst{5,6}
          \and
          Jairo Mendez-Abreu\inst{5,6}
          \and
          Justus Neumann\inst{15}
          \and
          Miguel Querejeta\inst{16}
          }

   \institute{European Southern Observatory, Karl-Schwarzschild-Str. 2, D-85748 Garching bei Muenchen, Germany
              \and Institute for Computational Cosmology, Department of Physics, Durham University, South Road, Durham DH1 3LE, UK
              \and Max-Planck-Institut f\"{u}r Astrophysik, Karl-Schwarzschild-Str. 1, 85748 Garching, Germany
              \and Centre for Extragalactic Astronomy, Department of Physics, Durham University, South Road, Durham DH1 3LE, UK
              \and Instituto de Astrofísica de Canarias, Calle Vía Láctea s/n, 38205 La Laguna, Tenerife, Spain
              \and Departamento de Astrofísica, Universidad de La Laguna, 38200 La Laguna, Tenerife, Spain
              \and Vyoma GmbH, Karl-Theodor-Str. 55, 80803 München
              \and Departamento de F\'{i}sica de la Tierra y Astrof\'{i}sica, Universidad Complutense de Madrid, E-28040 Madrid, Spain
              \and Instituto de F\'isica de Part\'iculas y del Cosmos (IPARCOS), Universidad Complutense de Madrid, E-28040 Madrid, Spain
              \and Department of Astrophysics, University of Vienna, Türkenschanzstraße 17, 1180 Wien, Austria
              \and Universidade de São Paulo, Instituto de Astronomia, Geofísica e Ciências Atmosféricas, Rua do Matão 1226, 05508-090, São Paulo, SP, Brazil
              \and European Space Agency, European Space Research and Technology Centre, Keplerlaan 1, NL-2200 AG Noordwijk, the Netherlands
              \and Department of Astronomy and Atmospheric Sciences, Kyungpook National University, Daegu, 41566, Republic of Korea.
              \and Astrophysics Research Institute, Liverpool John Moores University, IC2 Liverpool Science Park, 146 Brownlow Hill, L3 5RF Liverpool, UK
              \and  Institute of Cosmology and Gravitation, University of Portsmouth, Burnaby Road, Portsmouth PO1 3FX, UK
              \and Observatorio Astronómico Nacional, C/Alfonso XII 3, Madrid 28014, Spain
              }
    
   \date{Received September 15, 1996; accepted March 16, 1997}

% \abstract{}{}{}{}{} 
% 5 {} token are mandatory
 
  \abstract{The epoch in which galactic discs settle is a major benchmark to test models of galaxy formation and evolution but is as yet largely unknown. Once discs settle and become self-gravitating enough, stellar bars are able to form; therefore, determining the ages of bars can shed light on the epoch of disc settling, and on the onset of secular evolution. Nevertheless, until now, timing when the bar formed has proven challenging. In this work, we present a new methodology for obtaining the bar age, using the star formation history of nuclear discs. Nuclear discs are rotation-supported structures, built by gas pushed to the centre via bar-induced torques, and their formation is thus coincident with bar formation. In particular, we use integral field spectroscopic (IFS) data from the TIMER survey to disentangle the star formation history of the nuclear disc from that of the underlying main disc, which enables us to more accurately determine when the nuclear disc forms. We demonstrate the methodology on the galaxy NGC 1433 -- which we find to host an old bar that is  $8.0^{+1.6}_{-1.1}\rm{(sys)}^{+0.2}_{-0.5}\rm{(stat)}$ Gyr old -- and describe a number of tests carried out on both the observational data and numerical simulations. In addition, we present evidence that the nuclear disc of NGC\,1433 grows in accordance with an inside-out formation scenario. This methodology is applicable to high resolution IFS data of barred galaxies with nuclear discs, making it ideally suited for the TIMER survey sample. In the future we will thus be able to determine the bar age for a large sample of galaxies, shedding light on the epoch of disc settling and bar formation.}
  

   \keywords{galaxies:bulges -- galaxies:evolution -- galaxies:formation -- galaxies: kinematics and dynamics -- galaxies:stellar content -- galaxies:structure}

   \maketitle
%
%-------------------------------------------------------------------

\section{Introduction}
 
Constraining the processes that drive galaxy evolution in different cosmic epochs is still work in progress. At higher redshifts, external processes, such as mergers, galaxy interactions and gas inflows, have an important contribution on how galaxies evolve (e.g., \citealp{schreiber2006sinfoni}; \citealp{genzel2008rings}; \citealp{law2009kiloparsec}; \citealp{dekel2009formation}, \citealp{oser2010two}). As the Universe expands and interactions are less frequent, internal processes begin to play an important role in the evolution of galaxies (\citealp{kormendy2004secular}). 

Among the most important internal drivers of the evolution of disc galaxies are stellar bars, which efficiently redistribute angular momentum, as well as both stars and gas (e.g., \citealp{lynden1972generating}; \citealp{combes1985spiral}; \citealp{athanassoula2003angular}; \citealp{munoz2004inner}; \citealp{sheth2005secular}; \citealp{romerogomezetal2007}; \citealp{dimatteoetal2013};  \citealp{halleetal2015}; \citealt{fragkoudietal2016,fragkoudietal2017b}). Bars have been linked to a global quenching of star formation in galaxies (e.g., \citealp{masters2012galaxy}; \citealp{schawinski2014green}; \citealp{haywood2016milky}; \citealp{geron2021galaxy}) while also being responsible for inducing bursts of star formation in central regions (e.g., \citealp{ishizuki1990molecular}; \citealp{ellison2011impact}; \citealp{coelho2011bars}).{ Bars are common structures in the local Universe; indeed, many studies found that between $\sim 30\%$ -- $70\%$ of disc galaxies host bars (e.g., \citealp{eskridge2000frequency}; \citealp{menendez2007near}; \citealp{barazza2008bars}; \citealp{sheth2008evolution}; \citealp{aguerri2009population}; \citealp{nair2010fraction}; \citealp{buta2015classical}; \citealp{erwin2018dependence}). Therefore, understanding the time of their formation and how they affect their host galaxies is key to understanding the late-stage evolution of galaxies themselves. }

%As the bar ages, a number of its physical properties evolve. For example \cite{kim2015mass} showed that the light profile of the bar evolves from an exponential Sérsic index ($n_{\rm{bar}} \approx 1-2$) to a flat one ($n_{\rm{bar}} \approx 0.2$). In a study of 144 local galaxies from the Spitzer Survey of Stellar Structure in Galaxies sample (S$^4$G -- \citealp{sheth2010spitzer}), they found that the oldest bars should have $n_{\rm{bar}} \approx 0.2$. In addition, theoretical studies show that one can expect it to grow longer and stronger with its evolution (\citealp{athanassoula2013bar}). Other important observational characteristics that evolve with the bar are the bar-to-total light fraction (Bar/T), that increases as the bar ages, and the morphology, that evolves from an elliptical to a boxy shape (e.g., \citealp{kim2015mass}). In summary, studies that propose to time bar ages should be consistent with the expected observational evidence of it.

By investigating how galaxies evolve dynamically at different redshifts, studies find that a large fraction of galaxies at high redshift are rotationally supported (e.g., \citealp{shapiro2008kinemetry}; \citealp{schreiber2009sins}; \citealp{epinat2012massiv}; \citealp{wisnioski2015kmos3d}; \citealp{rizzo2020dynamically}; \citealp{lelli2021massive}), with fractions varying between $70-90\%$ at $z \sim1$ and $47-74\%$ at $z \sim 2$. Using JWST Early Release Observations of the galaxy cluster SMACS 0723, \cite{2022arXiv220709428F} show that the fraction of disc galaxies in the early Universe is still an open question. They find that disc galaxies dominate the morphology for $z\sim1.5$ with a factor of $\sim10$ higher then the previous results from HST. Nevertheless, these rotationally-supported discs at higher redshifts often present higher velocity dispersion, suggesting they are turbulent, unsettled and thick (e.g., \citealp{elmegreen2006observations}; \citealp{cresci2009sins}; \citealp{newman2013sins}; although see also \citealt{rizzo2020dynamically}). Since analytical and numerical work indicate that the bar can only form once the disc is dynamically settled, the moment of bar formation marks this transition epoch, in which the discs are at least partially dynamically cold (e.g. \citealp{kraljic2012two} and references therein). Exactly how and when the switch from external to internal processes as protagonists of galaxy evolution happens is still not clear, but it is associated to the settling of galactic discs. Therefore, timing the epoch of bar formation is a major step forward to piecing together the different phases of galaxy evolution.

% Since analytical and numerical work indicate that the bar can only form once the disc is dynamically settled, or at least part of it (see references in \citealp{gadotti2019time}), timing the epoch of bar formation is a major step forward that can bear the answer to the puzzle of when the disc has settled.
% Exactly how and when the switch from external to internal processes as protagonists of galaxy evolution happens is still not clear, but is clearly associated to the settling of galactic discs.

% Hence, timing the epoch of bar formation is a major step forward that can bear the answer to the puzzle of when the disc has settled.

Different studies have attempted to time when the bar formed for a handful of galaxies using different approaches. \cite{gadotti2015muse} analysed stellar populations in the nuclear disc of NGC\,4371 using high quality MUSE data, and found a bar age of about 10 Gyrs. \cite{perez2017observational} used the formation of the boxy/peanut-shaped bulge, originated from instabilities of the bar, to estimate the age of the bar in NGC\,6032 as 10 Gyrs. \cite{de2019clocking} analysed the star formation histories of nuclear structures including inner bars, and found that the inner bars in NGC\,1291 and NGC\,5850 must have formed at least $6.5$~Gyrs and $4.5$~Gyrs ago, respectively. These studies provide observational evidence that bars can be long-lived, which is in agreement with cosmological simulations that find bars forming between redshifts $1-2$ that survive down to $z=0$ (e.g. \citealt{Kraljicetal2012, rosasguevaraetal2020, Fragkoudietal2020, fragkoudietal2021}).

One of the immediate effects that follows bar formation -- and the ensuing onset of tangential forces in the disc -- is the gas inflow to the central parts of the galaxy. This gas inflow builds the rotation-supported structures known as nuclear discs -- commonly refereed to as pseudo-bulges or disc-like bulges. Simulations and theoretical studies find that, after the bar forms, it only takes $\sim 10^8$ yrs to form the nuclear disc (e.g., \citealt{athanassoula1992morphology,Athanassoula1992b}; \citealp{emsellem2015interplay}, \citealp{seo2019effects}; \citealp{baba2020age}). Thus, one can use the time of the formation of the nuclear disc -- obtained through galactic archaeology -- to time the age of the bar (\citealp{gadotti2015muse}). The study by \cite{gadotti2015muse} was offered as a proof of concept using NGC\,4371 to demonstrate the feasibility of timing bar formation with archaeological evidence from the nuclear disc itself. However, the nuclear disc in NGC\,4371 has no star formation and is mainly dominated by old stars, which makes it a special case for this type of analysis. More generally, with more complex star formation histories, estimating stellar population properties in the central region of disc galaxies is not a trivial task, since the observed light carries tangled information of the nuclear disc with the underlying main disc -- and possibly other structural components that were already present once the nuclear disc formed. Therefore, in order to reliably detect the oldest stars in the nuclear disc, we need a way to disentangle the light of the nuclear disc from the underlying main disc.

This is precisely the goal of this work, in which we present a new methodology for disentangling the light of the nuclear disc from other structures in the central region of the galaxy, enabling us to derive the bar formation epoch. With the disentangled light, we derive independent star formation histories (SFHs) for different stellar structures to time the moment the nuclear disc formed and, therefore, when the bar formed. We select NGC 1433, one of the galaxies in the TIMER survey (\citealp{gadotti2019time}) as a pilot galaxy with which to present this methodology, which will in the future be applied to all TIMER galaxies.

This paper is organised as follows. In Section~\ref{secDataDesc} we describe the data and motivate the selection of NGC\,1433 for this study. In Section~\ref{sec:DisLight} we describe the methodology that we develop to disentangle the SFH of the nuclear disc from that of the underlying main disc, and how this can be used to obtain the bar age, as well as the tests on the methodology carried out using hydrodynamic simulations. In Section~\ref{sec:results} we present our results on the age of the bar in NGC\,1433 and the detailed build up of its nuclear disc. In Section~\ref{secDisc} we discuss the implications of our results in the context of galaxy evolution and the formation of nuclear discs. We summarise and conclude in Section~\ref{sec:summary}. Additional tests on the methodology, including two control galaxies, are presented in the Appendix.

% %-------------------------------------------------------------------

\section{Sample and data description}
\label{secDataDesc}

\begin{figure*}
\centering
\includegraphics[width=0.7\linewidth]{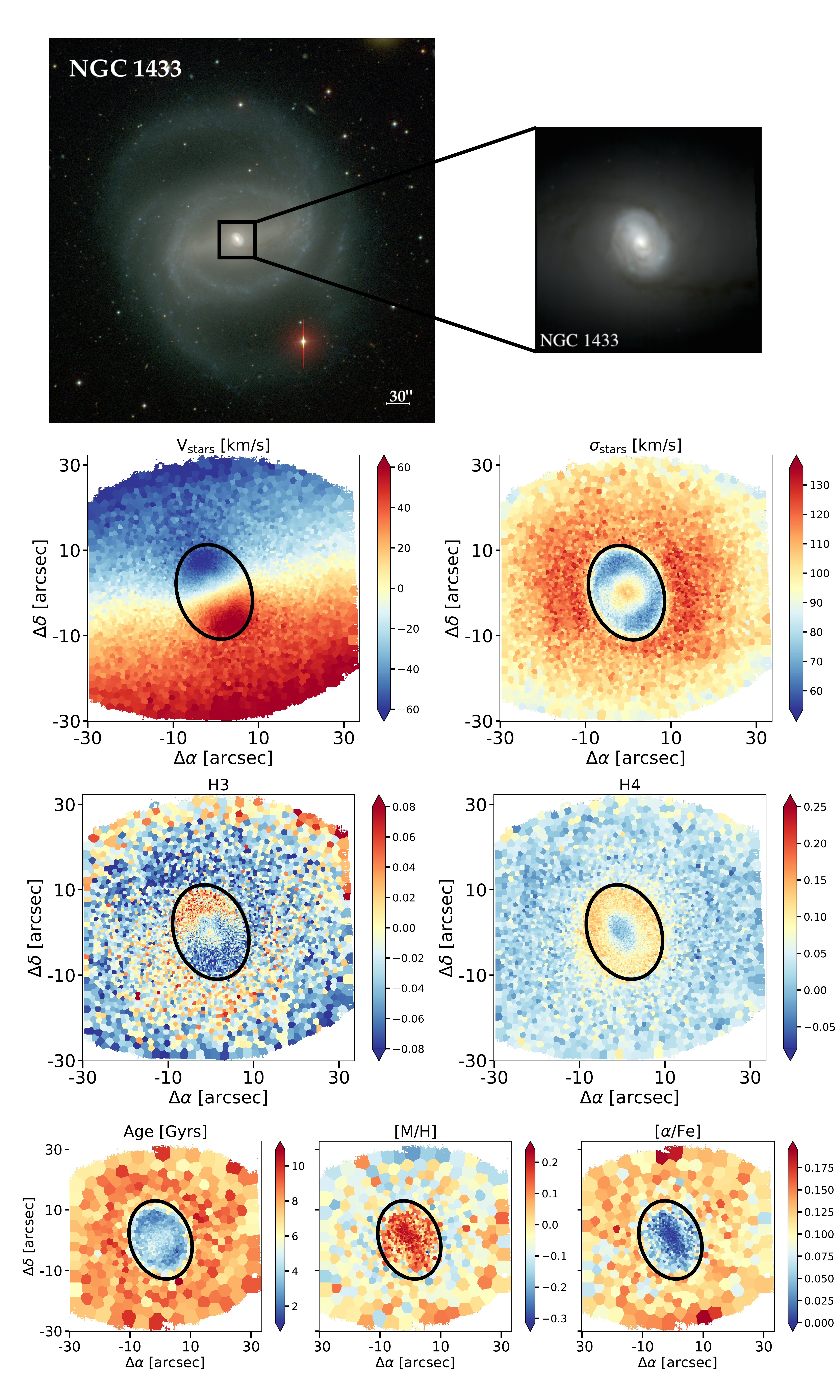}
    \caption{\textbf{Images}: colour composites of NGC\,1433 from the Carnegie-Irvine Galaxy Survey (top-left; \citealp{ho2011carnegie}) and TIMER (top-right; \citealp{gadotti2019time}). \textbf{Maps}: maps of stellar kinematics and population properties derived from the TIMER data using GIST: stellar velocity, velocity dispersion, h$_3$, h$_4$, age, metallicity and $\alpha$ enhancement. The nuclear disc radius is shown with a solid black contour, and displays a faster rotation, a drop in velocity dispersion, an anti-correlation between velocity and h$_3$ and an increase in h$_4$. In addition, the same region corresponds to a drop in mean age and $\alpha$ enhancement and an increase in metallicity. It is clear that NGC\,1433 hosts an younger nuclear disc with more rotational support in the central region than the underlying population, in agreement with \cite{gadotti2020kinematic} and \cite{bittner2020inside}.}
    \label{fig_DataDesc} 
\end{figure*}

% \begin{figure}
% \centering
% \includegraphics[width=0.7\linewidth]{plots/Section2.pdf}
%     \caption{\textbf{Images}: colour composites of NGC\,1433 from the Carnegie-Irvine Galaxy Survey (top-left; \citealp{ho2011carnegie}) and TIMER (top-right; \citealp{gadotti2019time}). \textbf{Maps}: maps of stellar kinematics and population properties derived from the TIMER data using GIST: stellar velocity, velocity dispersion, h$_3$, h$_4$, age, metallicity and $\alpha$ enhancement. The nuclear disc radius is shown with a solid black contour, and displays a faster rotation, a drop in velocity dispersion, an anti-correlation between velocity and h$_3$ and an increase in h$_4$. In addition, the same region corresponds to a drop in mean age and $\alpha$ enhancement and an increase in metallicity. It is clear that NGC\,1433 hosts an younger nuclear disc with more rotational support in the central region than the underlying population, in agreement with \cite{gadotti2020kinematic} and \cite{bittner2020inside}.}
%     \label{fig_DataDesc} 
% \end{figure}

In order to present the methodology we have developed -- which separates the light of nuclear discs from the underlying population -- we select NGC\,1433 for this pilot study, since it has a nuclear disc with similar properties to most nuclear discs in the TIMER sample (e.g., \citealp{gadotti2020kinematic}, \citealp{bittner2020inside} -- which will be future targets of study). Also, we ensure that our methodology does not produce artificial bar ages for unbarred galaxies by applying it to two control barless galaxies: NGC\,1380 and NGC\,1084 (see Appendix~\ref{appdx_control}).

\cite{buta2015classical} classified NGC\,1433 morphologically as strongly barred (SB) with a nuclear ring/lens (nrl) and a nuclear bar (nb). In addition, the galaxy is at a distance of 10 Mpc, has an inclination of 34$^\circ$, and stellar mass of $2\times10^{10}$ M$_\odot$ (see references in \citealp{gadotti2019time}). Following \cite{gadotti2020kinematic}, the radius of the nuclear disc is defined as the peak in the $v/\sigma$ radial profile, which is 440 pc (see Fig.~\ref{fig_VoS_Age}). We use data from the TIMER project: the observations were carried out using the MUSE instrument at the European Southern Observatory Very Large Telescope (ESO-VLT), in Period 97, from 2016 March to October, in Wide Field Mode. Considering the point spread function full width at half maximum, the spatial resolution of these observations is about 50 pc. Further details on the observations and data reduction can be found in \cite{gadotti2019time}. 

In Fig.~\ref{fig_DataDesc} we display colour composites of our target along with maps of stellar kinematics and population properties: stellar velocity and velocity dispersion, the higher order moments of the line of sight velocity distribution h$_3$ and h$_4$, stellar age, metallicity and $\alpha$ enhancement, which were produced applying the GIST pipeline (\citealp{bittner2019gist}) to the TIMER MUSE data (more details on the following Section). One can see that NGC\,1433 has clear signatures of a nuclear disc, that is, a rapidly-rotating structure detached from the main disc and coincident with a drop in velocity dispersion and mean ages. In addition, it also shows an anti-correlation between velocity and h$_3$, and an increase in h$_4$, as expected for nuclear discs. These results are all in agreement with the findings in \cite{gadotti2020kinematic} and \cite{bittner2020inside}, respectively.

% %-------------------------------------------------------------------

\section{Methodology}
\label{sec:DisLight}

\begin{figure*}
\centering
\includegraphics[width=\hsize]{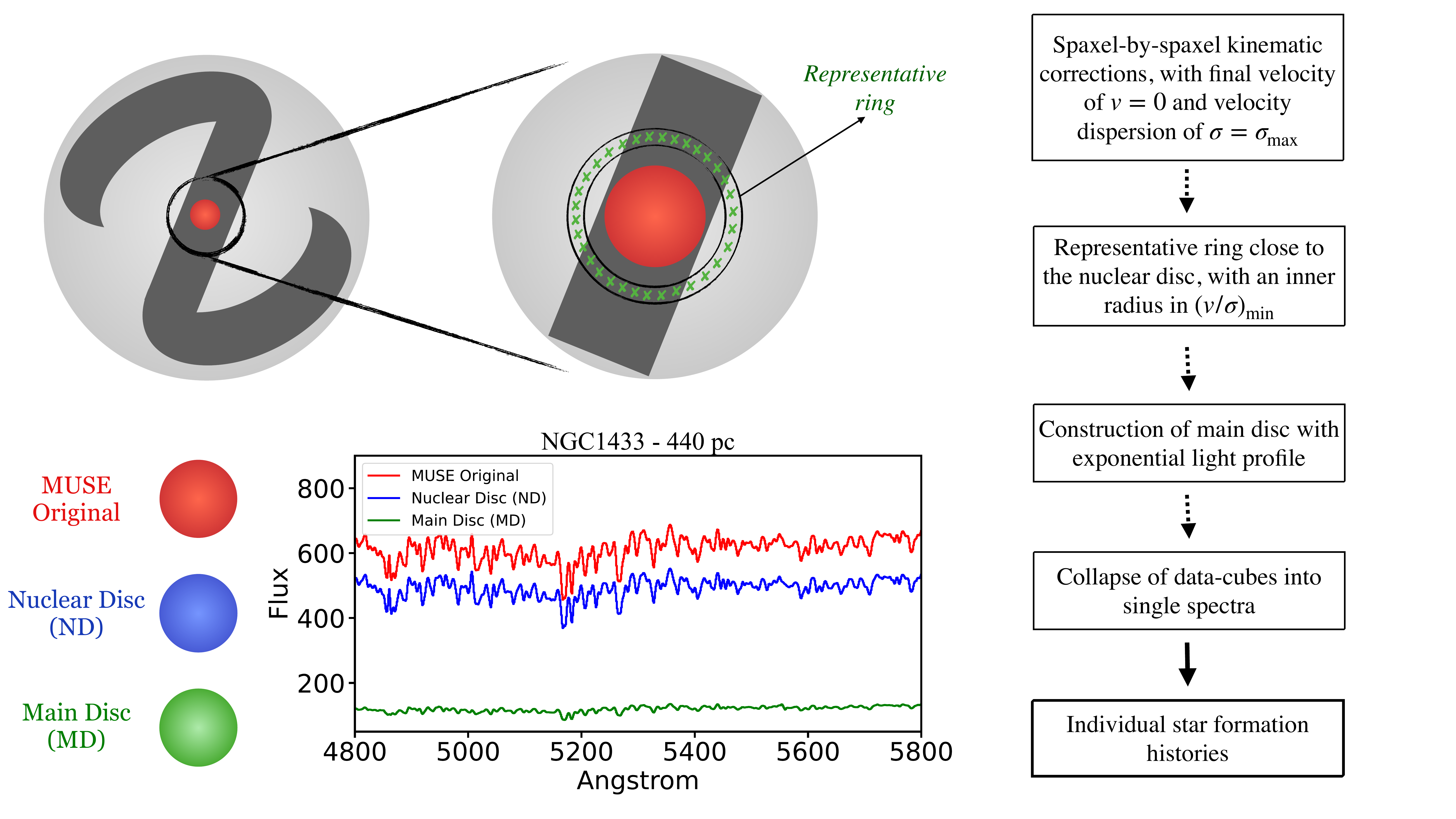}
    \caption{Illustration of the methodology described in Section~\ref{sec:DisLight} for a galaxy that hosts a nuclear disc. On the top-left, we highlight the field of view from MUSE in the centre of the galaxy, and the position of the representative region just outside the nuclear disc. The position of the representative ring is chosen taking into consideration the $v/\sigma$ radial profile. From it we derive the representative spectrum used to build the main disc. At the bottom, we display the output from the light disentangling: the original data cube (red), the nuclear disc data cube (blue), and the representative main disc data cube (green). All three data cubes have spaxels corrected to $v=0$ km/s and $\sigma = \sigma_{\mathrm{max}}$ km/s. Lastly, we collapse each data cube into a mean spectrum (as illustrated), and derive star formation histories for each one of those. The steps in our methodology are described on the right.}
    \label{fig_RepSpec}
\end{figure*}

In this Section we describe our methodology step-by-step, as illustrated in Fig.~\ref{fig_RepSpec}. We describe how we create the underlying stellar population contribution (Section~\ref{subsection3.1}) and how we isolate the star formation history of the nuclear disc (Section~\ref{subsection3.2}). In Section~\ref{subsection3.3} we describe the details in our data analysis and our criterion for obtaining the time of bar formation. In Section~\ref{subsection3.4} we test our methodology by applying it to a simulated barred galaxy.

\subsection{Building the underlying population contribution and disentangling the nuclear disc light}
\label{subsection3.1}

Recent studies deriving radial profiles of stellar ages found that galaxies hosting nuclear discs display a drop in average ages towards the centre. This drop in average ages is coincident with the radius of the peak in stellar $v$/$\sigma$ (e.g. \citealp{falcon2002bulges}, \citealp{bittner2020inside}), which implies that these galaxies have a central stellar structure that is younger than the main underlying population. This agrees with the scenario whereby nuclear discs are structures formed by a relatively late gas inflow induced by stellar bars (e.g, \citealp{gadotti2015muse}; \citealp{gadotti2019time}; \citealp{bittner2020inside}). Hereafter, we address as the "underlying main disc" every stellar population that was present before the formation of the nuclear disc. Since the observed light carries combined information of the younger nuclear disc and other underlying central structures, the ages derived for the central region of the galaxy can only be considered as an upper limit to the mean stellar age of the nuclear disc.  

In this section, we describe a new strategy to build the underlying main disc and disentangle the light of it from that of the nuclear disc. In summary, this methodology consists of deriving a spectrum that represents the main underlying disc (hereafter referred as \textit{representative spectrum}), using it to build a representative main disc datacube, and later subtract it from the observed data. We make two hypothesis: (1) the main disc extend all the way to the center, and its surface brightness profile follows an exponential law and (2) the stellar  populations of the main disc do not change significantly in the inner regions and, therefore,  we can assume is similar to the spectrum extracted from an  aperture surrounding the nuclear ring. The result from this subtraction is expected to be the isolated nuclear disc. A step-by-step outline and an illustration of our methodology are shown in Fig.~\ref{fig_RepSpec}. 

Before building the main disc, we treat the original datacube taking into consideration the kinematic properties. We first use the Data Analysis Pipeline from the PHANGS-MUSE survey (DAP – \citealp{emsellem2021phangs}) for the original datacube, deriving kinematic and line emission properties. DAP is a module-based pipeline based on the Galaxy IFU Spectroscopy Tool (GIST; \citealp{bittner2019gist}) able to extract properties from datacubes such as kinematic information, emission lines fluxes and more. For the emission lines fitting, DAP uses \texttt{pPXF} (\citealp{cappellari2012ppxf}), considering emission lines as extra gaussian elements added to the stellar continuum. We measured the kinematic maps for Voronoi-binned (\citealp{cappellari2003adaptive}) data with signal to noise of 100 and emission lines spaxel-by-spaxel, between the wavelengths 4800 and 7000 \r{A}, employing the E-MILES simple stellar population (SSP) model library (\citealp{vazdekis2016uv}) to remove the stellar continuum. From the results, we have kinematic properties such as stellar velocity and velocity dispersion and emission line fluxes. Applying the kinematic information, we shift all spectra to have a final velocity of $v=0$ km/s, accounting for both the galaxy recession velocity and internal kinematics. In addition, we also convolve the original data to ensure that every spectra has $\sigma=\sigma_{\mathrm{max}}$, where the latter is the highest velocity dispersion in the central region of the original data cube. For NGC\,1433, $\sigma_{\mathrm{max}}$ is $122$ km/s. We point out that we verified that the convolved data and the original data result in the same stellar population, so such procedure does not affect our results. Our aim in doing so is to ensure we do not create artificial wings or artificial emission/absorption lines when subtracting the spectra (see below). Once the original data is shifted to $v=0$ km/s and $\sigma=\sigma_{\mathrm{max}}$ km/s, we build our underlying main disc.

\begin{figure}
\centering
\includegraphics[width=\linewidth]{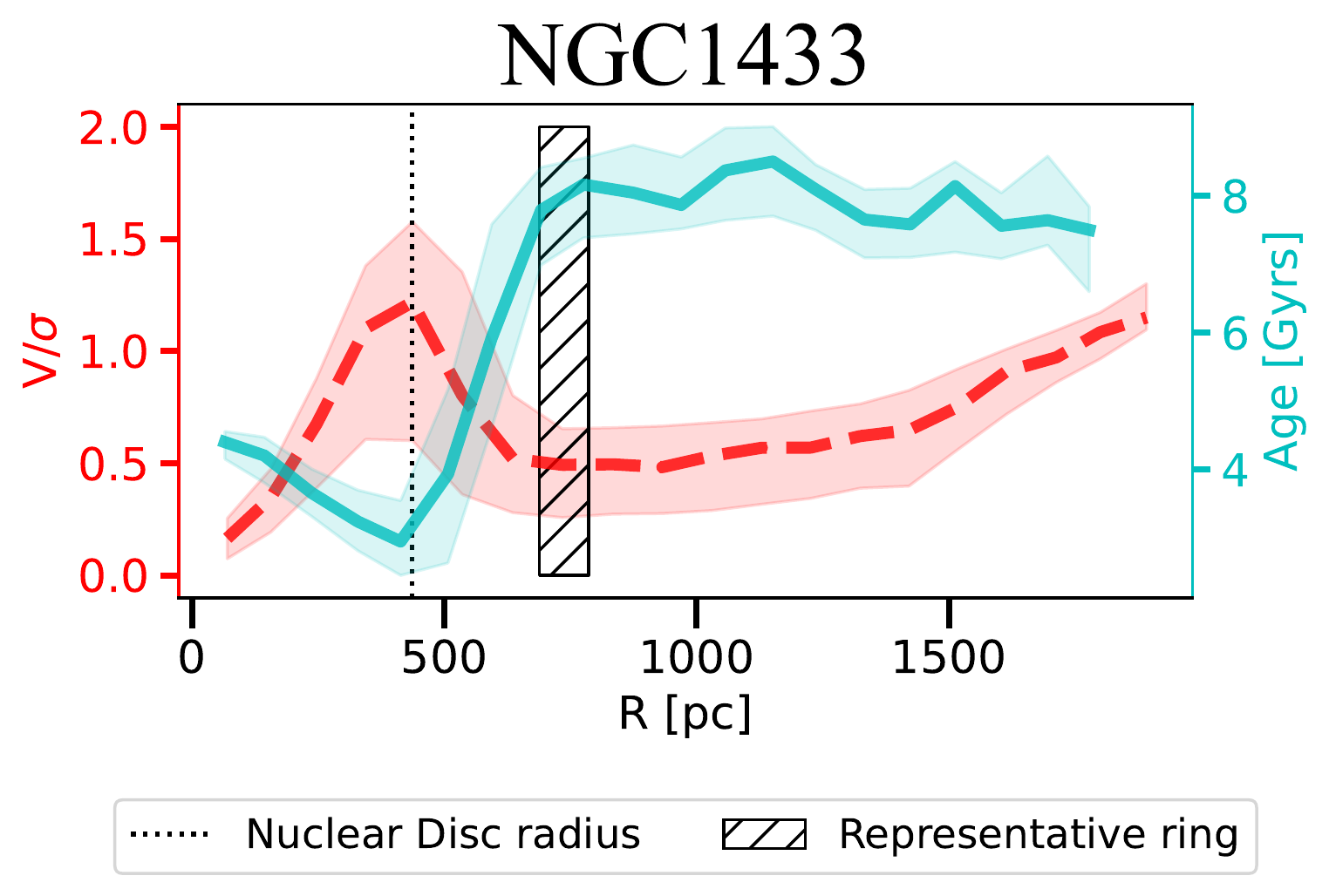}
    \caption{$v/\sigma$ radial profile displayed in red-dashed contours (values in the left-axis) and light-weighted average age in solid-blue contours (values in the right-axis) for NGC\,1433. We display the median values together with the 1st and 3rd quartiles. The dotted-black vertical line marks the nuclear disc radius, and the hatched area the representative ring. Note that the representative ring is placed in the first $v/\sigma$ minimum outside the nuclear disc, which matches the oldest mean age.}
    \label{fig_VoS_Age}
\end{figure}

The first step is to define the representative region from which we will extract the representative spectrum. This region is chosen as a ring surrounding the nuclear disc -- from now on referred to as the \textit{representative ring}. The representative ring is expected to be as close as possible to the nuclear disc without being contaminated by its light. For galaxies hosting a nuclear disc, such as NGC\,1433, we take into consideration the stellar $v/\sigma$ radial profile (Fig.~\ref{fig_VoS_Age}). We define the inner radius of the representative ring as the first minimum point outside the nuclear disc. That is the radius outside which we expect the light from the main disc to start dominating and, as expected, it is also the radius with the oldest mean age (see Fig.~\ref{fig_VoS_Age}). Nevertheless, we tested different positions for the representative ring to assess the systematic error linked to this decision (see Appendix~\ref{apdx_ring}). In addition, the representative ring has a width of 2’’, so one does not expect strong age and flux gradients inside of it. Nonetheless, we performed tests with widths between 1’’ and 4’’ with little variation in the outcome. 

\begin{figure*}
\centering
\includegraphics[width=0.9\linewidth]{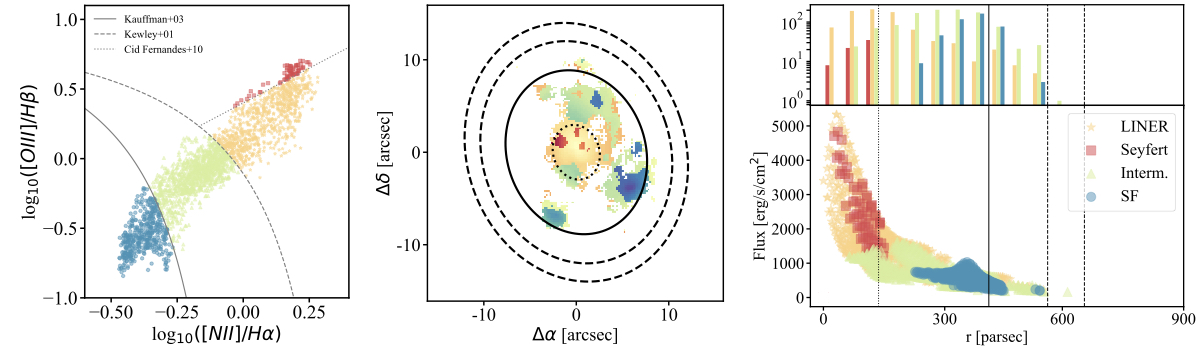}
    \caption{\textbf{Left:} BPT diagram (\citealp{baldwin1981classification}) classification of each spaxel for the nuclear disc in  NGC\,1433. The BPT classification is carried onto the \textbf{middle} and \textbf{right} panels, that display the physical position of the different spaxels and the light radial profile, respectively. In addition, we also display the radius of the inner mask dominated by AGN contribution (dotted contours), the radius of the nuclear disc (solid contours) and the representative ring (dashed contours).}
    \label{fig_BPT}
\end{figure*}

Secondly, we aim to mask spaxels dominated by AGN emission, based on the Baldwin, Phillips \& Tervelich diagrams classification (BPT diagrams; \citealp{baldwin1981classification}). We use the emission lines H$\alpha$, H$\beta$, [OIII], and [NII] extracted by DAP, as described above, to build the BPT diagram in Fig.~\ref{fig_BPT}.%, using a threshold of amplitude over noise above 10 to select emission lines. 

In the third step, we normalize the flux in each spaxel of the representative ring to account for the radial flux increase in the main disc towards the centre. We normalize each spectra to $r=0$ pc, assuming that the disc light profile follows an exponential function:

\begin{equation}
    I(r) = I_0 \times e^{-r/h}
    \label{eq:Iexp}
\end{equation}

\noindent where $h$ is the disc scale-length, $r$ is the distance of a given spaxel from the centre (corrected for inclination effects), and $I_0$ is the flux at $r=0$ pc. The value of $h$ is taken for NGC\,1433 comes from \cite{salo2015spitzer}, and is 3100 pc. Considering we have $I(r)$ for each spaxel in the representative ring, we divide it by $e^{-r/h}$, extrapolating the observed flux to the centre of the galaxy. We derive the representative spectrum as the mean flux per wavelength of all spectra from the representative ring with $v=0$ km/s, $\sigma = \sigma_{\mathrm{max}}$ km/s and the corresponding flux at $r=0$ pc. Finally, to re-construct a data cube of the main disc, we extrapolate the representative spectrum back to a range of radii, taking into account again an exponential light profile for the main disc. Although the main disc is often described with an exponential light profile, recent studies show that discs may not follow an exponential light profile all the way to the centre (e.g., \citealp{zhu2018morphology}; \citealp{breda2020indications}; \citealp{papaderos2021inside}). In light of this, we tested the implications of assuming an exponential profile in Appendix~\ref{apdx_prof}, by applying the same methodology to a flat light profile main disc. We show that the choice of a profile for the main underlying population does not affect the results noticeably.

\begin{figure}
    \centering
    \includegraphics[width=\linewidth]{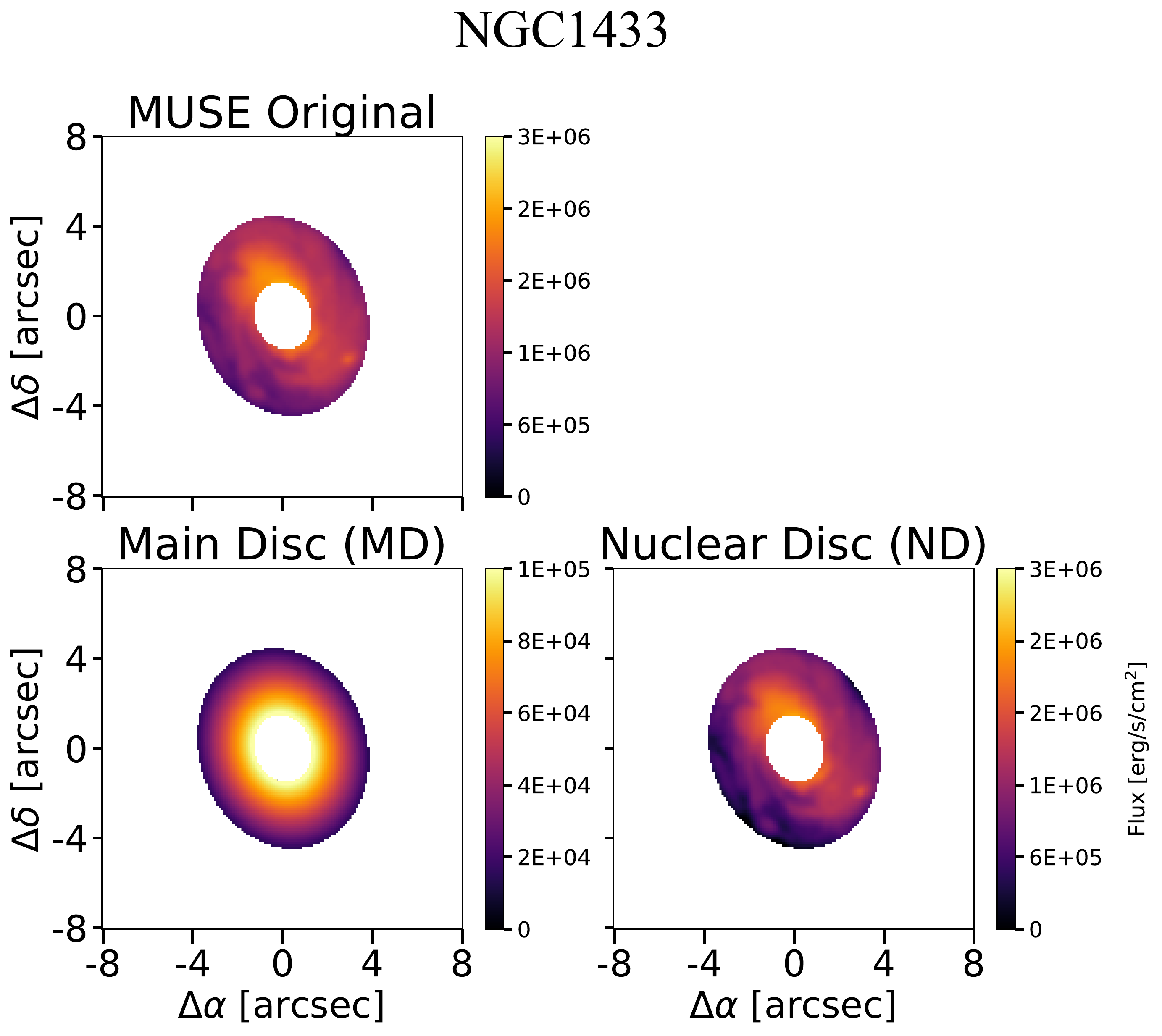}
    \caption{Illustration of different data products derived from the methodology described in Sect.~\ref{sec:DisLight}. From left to right, we display the sum of MUSE fluxes between 4800 and 5800 \r{A}, of the original data cube, the derived representative main disc and the nuclear disc data, which is the result of subtracting the representative main disc from the original data cube. For the representative main disc, it is possible to notice the exponential increase of flux towards the centre.}
    \label{fig_datacubes}
\end{figure}

Next, we subtract the reconstructed main disc from the original data. As mentioned, in order to prevent the creation of artificial wings from the subtraction, we also use the original data convolved and shifted to $v=0$ km/s and $\sigma = \sigma_{\mathrm{max}}$ km/s. With that, we can disentangle our original data into the Main Disc (MD) and the Nuclear Disc (ND) data cubes, as exemplified in Fig.~\ref{fig_datacubes}. Lastly, we collapse each data cube into a single mean spectra, deriving star formation histories (SFH) for each one. To assure that the collapsed spectra is the correct description of the central region for the three data cubes, we masked contributions from AGN in the center. Using the BPT classification together with the light radial profile (Fig.~\ref{fig_BPT}, right), we delimited a central region to be masked of around $140$ pc (15 spaxels), since it is dominated by AGN emission and can contaminate the total flux of the nuclear disc. 

Due to the high quality of the TIMER data (\citealp{gadotti2019time}), the decision to collapse the datacube is not motivated by the increase of signal-to-noise ratio. Instead, we collapse the datacube to save computational time and guarantee low statistical errors. Nevertheless, we applied the same methodology for the non-collapsed datacube and the bar age is not strongly affected (see Appendix~\ref{app_noncollapsed}).
 
\begin{figure*}
    \centering
    \includegraphics[width=\linewidth]{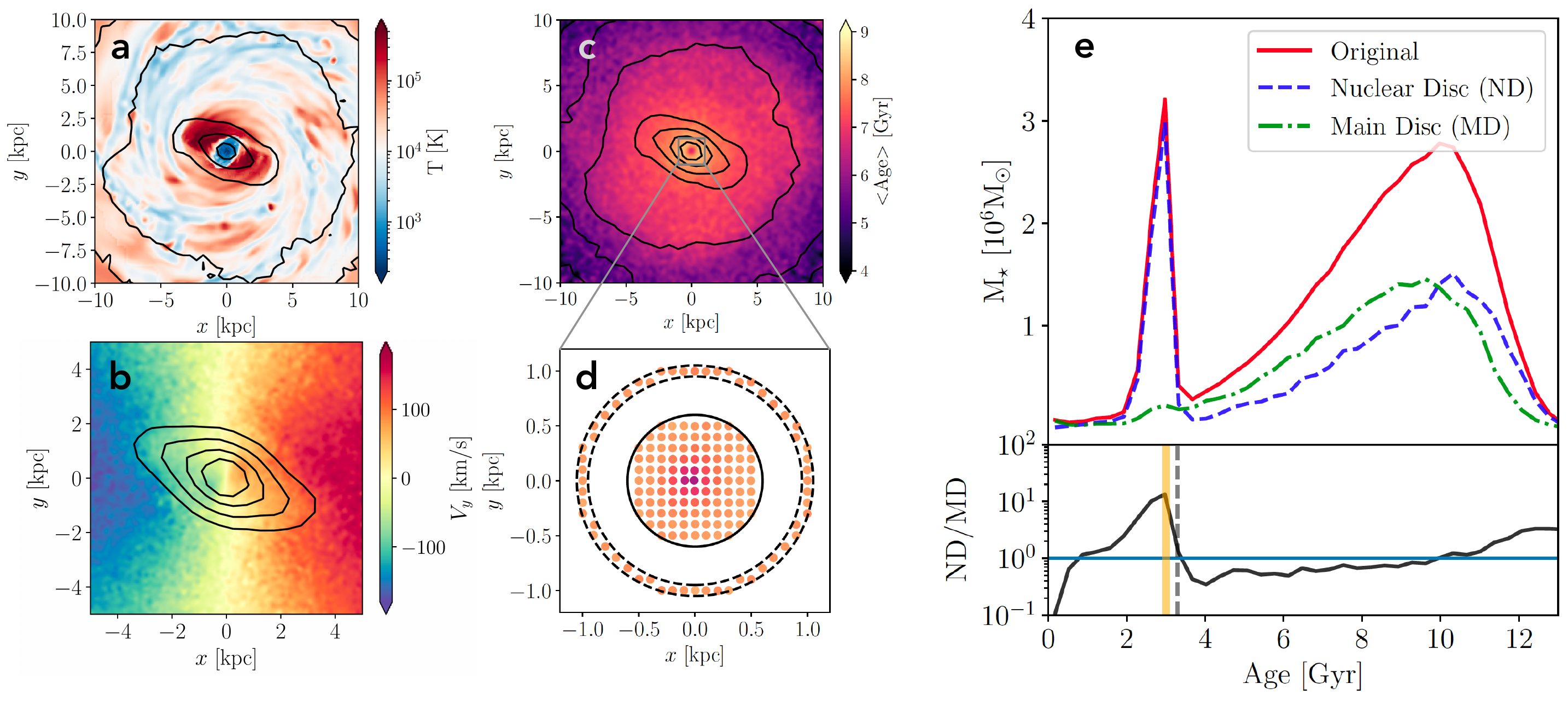}
    \caption{{\bf{Testing the methodology using hydrodynamic simulations:}} \emph{\bf{a:}} Face-on projection of the gas temperature. The dust lanes on the leading edge of the bar and the gaseous nuclear disc are clearly visible as low temperature regions.
    \emph{\bf{b:}} Kinematic map showing the velocity of stars in the $y$ direction with stellar isodensity contours overplotted, which outline the shape of the bar (note the different scale from panel \emph{a}). A highly rotating stellar component in the central kiloparsec -- i.e. the nuclear disc -- is clearly evident in the kinematic map.
    \emph{\bf{c:}} Age map of all the stars in the simulated galaxy. The gray box shows the inset which is represented in panel \emph{d}.
    \emph{\bf{d:}} A zoom in around the nuclear disc region. The inner and outer dashed lines denote the region used to obtain the SFH of the 'representative disc region'. The solid black line denotes the radius within which the SFH of the nuclear disc is estimated. The scatter points indicate the locations of the pixels used for deriving the SFHs in panel \emph{e}. The points are colour-coded by the mean age in the pixel. \emph{\bf{e:}} The top panel shows the SFHs of the original nuclear disc region (solid red), of the representative SFH of the main disc (dot-dashed green), and of the nuclear disc with the representative SFH subtracted (dashed blue). The bottom panel shows the ratio of the subtracted SFH to the representative SFH. The vertical orange line indicates the time of bar formation in the simulation (3\,Gyr), and the vertical dashed line indicates the time at which the ND/MD is above 1 (3.3\,Gyr).}
    \label{fig_sims}
\end{figure*}

\subsection{Deriving Star Formation Histories (SFHs) and mass assembly}
\label{subsection3.2}

For each collapsed spectra -- MUSE original, Main Disc (MD), and clean Nuclear Disc (ND) data --, we run the GIST pipeline (\citealp{bittner2019gist}) to derive stellar population properties and star formation histories. To guarantee consistency in our analysis with previous TIMER work, we use the same GIST configuration as in \citet{bittner2020inside}. Firstly, GIST employs an unregularised run of \texttt{pPXF} (\citealp{cappellari2004parametric}; \citealp{cappellari2017improving}) to derive stellar kinematic properties. We also include a low order multiplicative Legendre polynomial in the fit to account for small differences between the shape of the continuum templates and the observed spectra. Next, GIST employs \texttt{pyGandALF} (see \citealp{bittner2019gist}) to model emission lines as Gaussians, simultaneously fitting the stellar continuum while the stellar kinematics remains fixed from the previous step. With this, we obtain the emission-subtracted spectra. Lastly, GIST performs a regularized \texttt{pPXF} run in the emission-subtracted spectra, in order to fit a combination of stellar populations and derive mean properties. Since metallicity and stellar velocity dispersion can both be responsible for absorption line broadening, causing possible degeneracies (e.g., \citealp{sanchez2011star}), we keep the stellar kinematics fixed from previous steps. In addition, to account for extinction and continuum mismatch effects, we apply an 8th order multiplicative Legendre polynomial in the fit. For the last step, we employ the MILES library (\citealp{vazdekis2015evolutionary}), light-weighted, with [M/Fe] between $-1$ and $+0.4$, ages in the range $0.03$--$14$ Gyr, and [$\alpha$/Fe] enhancements of $+0.0$ and $+0.4$. Lastly, we use the regularization error value of 0.15 derived for TIMER data by \cite{bittner2020inside}. As described in \cite{cappellari2017improving}, the regularization of the star formation histories allows one to derive the smoothest and still physically meaningful result. To assess how much our final bar age relies on the regularization error, we tested different values in Appendix~\ref{appdx_regul}. For further details on the data analyses, we refer the reader to previous TIMER papers (e.g., \citealp{bittner2020inside}). 

\texttt{pPXF} also estimates different weights for single stellar populations (SSPs) with different ages, allowing us to build a non-parametric star formation history (SFH). Each weight represents the fraction of the light formed in the different SSPs. In order to convert the light-weighted SFHs to mass-weighted SFHs, we consider the distance to the galaxy to derive the intrinsic luminosity. We then use the mass-to-light ratios\footnote{http://research.iac.es/proyecto/miles/pages/predicted-masses-and-photometric-observables-based-on-photometric-libraries.php} predicted from the BaSTI isochrones (\citealp{pietrinferni2004large}, \citeyear{pietrinferni2006large}, \citeyear{pietrinferni2009large}, \citeyear{pietrinferni2013basti}) to convert luminosity into stellar mass. The mass-to-light ratios assume a Kroupa revised IMF (\citealp{kroupa2001variation}), and the MILES template library (\citealp{vazdekis2015evolutionary}) with [$\alpha$/Fe] enhancements of $+0.0$ and $+0.4$. These mass-to-light ratios depend on the age, metallicity and [$\alpha$/Fe] enhancement that best describe the observed data, and account for both stellar and remnants masses. 

\subsection{Age-dating bar formation}
\label{subsection3.3}

Considering the scenario in which the nuclear disc is formed from a late gas inflow due to the bar, in an ideal case, one could expect the original SFH characterised by an older star formation event followed by a later burst. Once we are able to disentangle the main disc and the nuclear disc, we can expect that the former's SFH would map the oldest star formation event, and the latter's SFH would map the youngest burst. 

If we could perfectly disentangle the light of the nuclear disc from the underlying main disc, one could simply use the first peak in the nuclear disc SFH as the formation time of the nuclear disc and, consequently, the bar. However, due to gradients in the stellar population properties, the region around the nuclear disc which we use to obtain the SFH of the main disc, might not be identical to the real SFH of the main underlying disc within the nuclear disc region. Thus, it might not fully remove the contamination of the main disc from the nuclear disc light.
With that in mind, and by testing our methodology on N-body+hydrodynamic simulations (see Section \ref{subsection3.4}), we employ a criterion to time the bar formation epoch as \textit{the moment when the nuclear disc dominates the star formation}, as a signature of the bar bringing gas towards the centre. {That corresponds to the first time in which the ratio between the star formation in the nuclear disc and that in the main disc, ND/MD, rises above 1, with a positive slope towards younger ages.} In order to verify whether this is a reliable criterion, we test our methodology using an N-body+hydrodynamic simulation of a barred galaxy below.

% In order to verify whether this is a reliable criterion, we apply our methodology to simulated galaxies and assess its robustness and limitations in Section~\ref{subsection3.4}. %For the simulated galaxies -- further discussed in the following section -- one can notice the limitations of this method. %The main caveat is that our methodology cannot fully capture the very oldest ages in the underlying main disc, due to the age gradient in the underlying main population. Thus, the SFH of the subtracted nuclear disc can still show remnants of old star formation which are not captured in the representative spectrum. In this case (see Fig. 6), the old peak present in the nuclear disc SFH is older than the representative spectrum SFH -- although the difference (of the order of 2 Gyr) is relatively small considering the uncertainties in estimating stellar ages around 10 Gyr. That is, only the very oldest star formation is not mapped. Therefore, the ND/MD ratio does go below 1 at old ages and does come back above 1 when the bar is formed. In conclusion, the simulated galaxies show that, even with this limitation, the chosen criterion (ND/MD$>1$) is still effectively tracing the epoch of the bar formation. We also point out that, given that the radius from which the representative spectrum is extracted is typically of the order of only a few hundred parsecs, effects from the presence of any age gradient should generally be small.

\subsubsection{Testing the method using hydrodynamic simulations}
\label{subsection3.4}

In order to test the robustness of the methodology developed here and some of the assumptions employed, we use an N-body and hydrodynamic simulation of an isolated Milky Way-like disc galaxy, which forms a bar and a nuclear disc self-consistently (the simulation is part of a suite of models developed to study the evolution of barred galaxies; Fragkoudi \& Bieri, in prep.). The simulation has two collisionless components (a stellar disc and a dark matter halo) and a collisional component (gaseous disc), which is able to form stars which subsequently return mass, energy and metals to the ISM via supernova feedback. We refer the reader to Appendix~\ref{sec:simtech} for technical details of the simulation. 

The simulation is evolved for a total of 3.3\,Gyr. The axisymmetric stellar disc which is in place at the start of the simulation (which we refer to as the `old stellar component'), rapidly forms a bar after $\sim$0.3\,Gyr (we define the bar as being fully formed when the $m=2$ Fourier mode of the surface density, $A_2$>0.3). During and after the formation of the bar, gas piles up at the leading edges of the bar, where it shocks, loses angular momentum, and is funneled to the centre, where it forms a dense gaseous nuclear disc (see panel \textbf{(a)} of Figure~\ref{fig_sims}), which proceeds to form stars. These new stars -- formed out of gas pushed to the centre by the bar -- form a highly rotating stellar nuclear disc (see panel \textbf{(b)} of Figure~\ref{fig_sims}), whose size is set by the bar orbits in the inner regions (see \citealt{Athanassoula1992b}), similar to those observed in the local barred galaxies (e.g. \citealp{gadotti2020kinematic}). 

We highlight that, while the new stars formed out of the gas have self-consistent ages -- according to when they are formed in the simulation, the 'old stellar component' can have any star formation history we assign to it.
In order to model the age gradient often found in galaxies, we assign ages to the old stellar component at the initial snapshot, such that this gives rise to a negative age gradient\footnote{In practice this is done by sampling from Gaussian distributions at each radius, with a decreasing mean value for the age.}, i.e. with older stars in the centre and younger stars at the edge of the disc.  This old underlying population, together with the new stars formed out of the gas in the simulation, give rise to the age map of the galaxy at the end of the simulation shown in panel \textbf{(c)} of Figure~\ref{fig_sims}. 

We can now extract the star formation history for a given `pixel' in the nuclear disc region -- as we do in the observations -- which will contain stars born from gas pushed to the centre by the bar, as well as old stars which were present before the bar formed. We also extract the star formation history for pixels in the region just outside the nuclear disc, which give us the representative SFH of the underlying main stellar disc (MD). Therefore, as in the methodology used for the observations, we can extract the SFH of both the nuclear disc region and the main disc. We then subtract the SFH of the main disc from the SFH in the nuclear disc region, in order to obtain the SFH of the "clean" nuclear disc itself (ND). These SFHs are shown in panel \textbf{(e)} of Figure~\ref{fig_sims}, with solid red for the original total SFH within the nuclear disc region, dot-dashed green for the SFH of the main underlying disc, and dashed blue lines for the "clean" nuclear disc SFH. {This can be compared to the ``true'' SFHs of the ND and MD, which are shown in the left panel of Fig. \ref{fig:agegradienttests} in Appendix C.} 

We find that due to the age gradient in the underlying disc as well as the gradient in the light profile, there can be a contamination of the `old component' in the subtracted ND SFH, which cannot be fully removed by subtracting the underlying main disc (see Appendix \ref{sec:simtech}, Figure \ref{fig:agegradienttests}, where we show how assuming a negative age gradient versus a flat age gradient affects the methodology). The extent to which there is `contamination' by the oldest stars in the ND will depend on both how steep the age gradient is, and on the location of the ring used to obtain the SFH of the main disc (see also Appendix \ref{sec:simtech}, Figure \ref{fig:ringloctest}). This indicates that we cannot simply use the oldest peak in the subtracted SFH (ND) in order to obtain the time at which the nuclear disc formed, but we rather should use the comparison between the subtracted nuclear disc and the representative SFH of the main disc (i.e. ND/MD). In practice, this is the time when the SFH of the nuclear disc increases above that of the main disc, i.e. when ND/MD rises above 1, with a {positive slope towards younger ages}. 

As can be seen from panel \textbf{(e)} of Figure~\ref{fig_sims}, this method allows us to recover the time at which the bar formed (which in the simulation occurs at $t_{\rm lookback}\sim$3\,Gyr), which is marked with the vertical solid orange line, while the ratio of the nuclear disc over the representative main disc SFH (ND/MD) gives a bar age of $t_{\rm lookback}$=3.3\,Gyr. Therefore we find that, even with the contamination of older ages in the subtracted spectrum, we can recover the time of bar formation, as the first time at which ND/MD rises above one, with an accuracy which will depend on the width of the age bins in the SFHs.
The main limitation in obtaining the bar age using this methodology therefore stems from uncertainties in deriving stellar ages -- and therefore the SFH -- which are typically of the order of $\sim1$\,Gyr (see e.g. \citealt{bittner2020inside}).

\section{Results}
\label{sec:results}

\begin{figure*}
\centering
\includegraphics[width=0.7\linewidth]{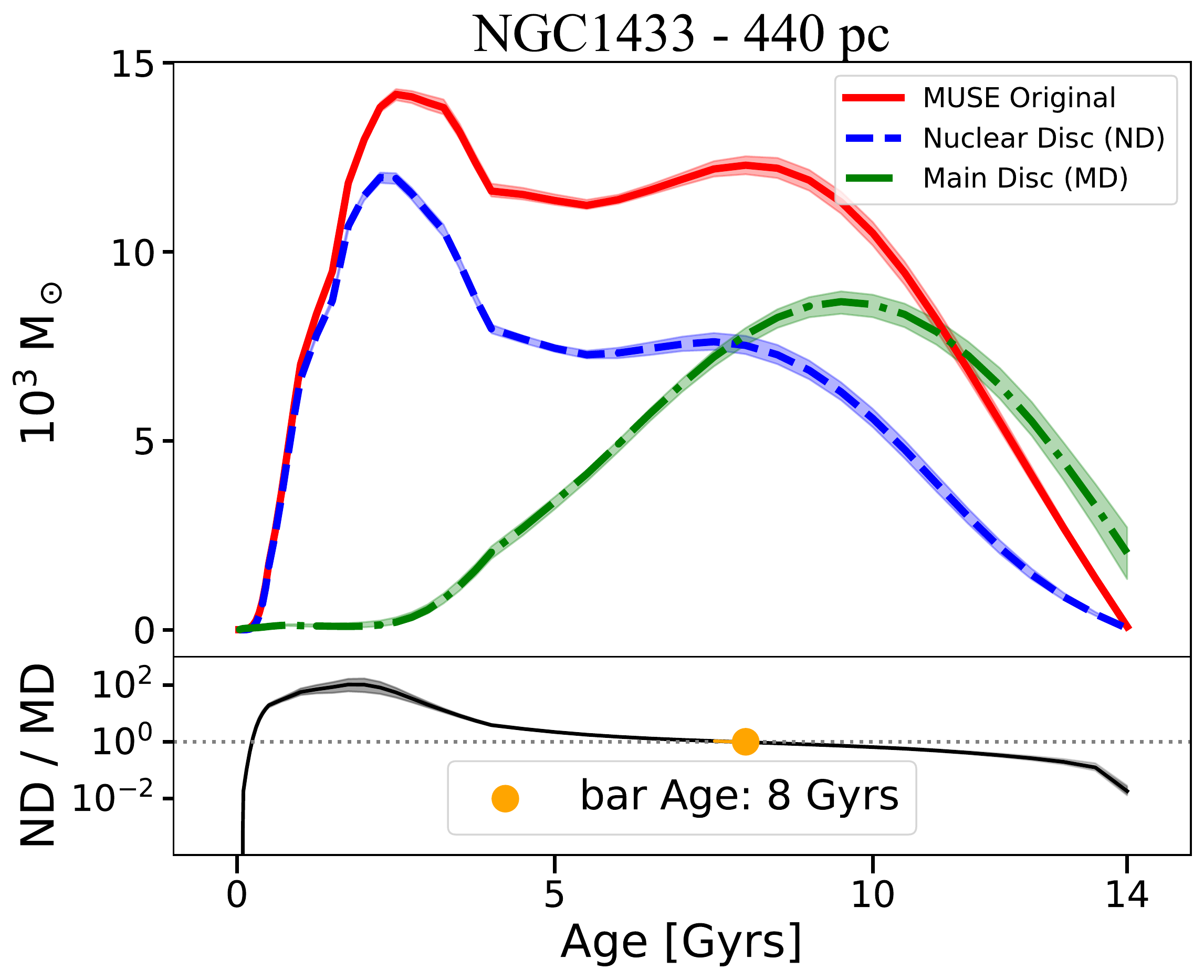}
    \caption{Mass-weighted star formation histories
    for NGC\,1433 from collapsed spectra of the MUSE original data, nuclear disc data and representative main disc data, as illustrated in Fig.~\ref{fig_RepSpec}. The y-axis corresponds to the mass formed in solar masses for each age bin and the x-axis to the age in Gyr, that is, the look-back time. The original data is in solid-red line, the subtracted data is in dashed-blue line and the representative spectrum is in dot-dashed-green line. The lines are the SFHs derived from the data while the shaded regions are results from 100 Monte Carlo runs. In the bottom panel we display ND/MD in black, and highlighted with an orange circle is the age where ND/MD is for the first time rising above one, together with the statistical uncertainty, that is $^{+0.2}_{-0.5}(\rm{stat})$ Gyr.}
    \label{fig_SFH}
\end{figure*}

After following the methodology to disentangle SFHs described in Section~\ref{sec:DisLight}, we present our results for our pilot study galaxy, NGC\,1433. We remind the reader that the chosen criterion to time bar formation epoch is the first moment when the ratio ND/MD increases above 1 with {positive slope towards younger ages}.

Figure~\ref{fig_SFH} shows our main results: the mass of stars formed with different age bins of the stellar templates -- analogous to SFHs -- for the MUSE original (solid-red line and contour), the main disc (dot dashed green line) and the nuclear disc (dashed blue line and contour) for NGC\,1433, together with the ratio ND/MD in the bottom panel. The age is related to the SSP template combination that best fits the observed spectra and can be understood as a “look back time”, i.e., $t=0$ is the present time. 

We measure the bar formation epoch of NGC\,1433 to occur $8.0^{+1.6}_{-1.1}\rm{(sys)}^{+0.2}_{-0.5}\rm{(stat)}$ Gyrs ago, corresponding to a redshift of $z \approx 1$. In order to quantify the statistical error of the methodology, we perform 100 Monte Carlo runs for each of the collapsed spectra (original data, nuclear disc data and representative spectra) to derive variations on the bar age. We use the noise information to sample a distribution of fluxes for each wavelength, creating 100 artificial spectra. We then run pPXF on each of these spectra, to obtain the different SFHs and the subsequent bar ages. This is shown in Fig.~\ref{fig_SFH} as the shaded area of each SFH. From it, we derive a statistical uncertainty corresponding to $^{+0.2}_{-0.5}\rm{(stat)}$ Gyr. The statistical error is subdominant, since, by collapsing the data cubes into single spectra, we achieve signal-to-noise values over 2000.  To further quantify uncertainties in the derived bar age, which can be introduced due to various aspects of the methodology, we perform multiple tests with different configurations (see Appendix~\ref{app_systematic}): different locations for the representative ring, different light profiles to describe the increase in density of the main underlying population towards the centre, and different regularization errors for the \texttt{pPXF} run that results in the derivation of the SFHs. From these tests, we find that there is a systematic uncertainty in our measurements of the bar age of the order of $^{+1.6}_{-1.1}\rm{(sys)}$ Gyr, which we quote in addition to our statistical errors.

From the SFHs, we can derive estimates of the total stellar mass in the isolated nuclear disc and the underlying main disc within the nuclear disc radius, by summing the mass formed through time following each curve. However, if NGC 1433 has significant age gradients within the central kpc, the very oldest population in the underlying disc may still be partly present in the isolated nuclear disc, as discussed above. Therefore, these would be, respectively, an upper limit to the mass of the nuclear disc, and a lower limit to the mass of the underlying population within the nuclear disc region. For NGC\,1433, we measure the underlying main disc mass within the nuclear disc radius as $2.95 \times 10^{8}$ M$_\odot$ and the `cleaned' nuclear disc mass as $4.05 \times 10^{8}$ M$_\odot$. To explore whether these values are in agreement with the literature, we extrapolate the underlying main disc mass to obtain the total mass of the galaxy assuming an exponential function, following Eq.~\ref{eq:totMass}:

\begin{equation}
    M_h = 2\pi \int_0^\infty{\Sigma (r) r dr} = 2\pi \Sigma_0 h^2, 
\label{eq:totMass}    
\end{equation}

\noindent where $\Sigma_0$ is the mass density at the centre and $h$ is the disc scale-length. Using the mass of the underlying main disc within the nuclear disc radius, we measure the mass density as $512$ M$_\odot$/pc$^{2}$. This value gives the extrapolated mass for the entire galaxy of $2.74 \times 10^{10}$ M$_\odot$, which is consistent with the total stellar mass of $2 \times 10^{10}$ M$_\odot$ derived by \citet{munoz2015spitzer}, considering the uncertainties involved. This indicates that our measurements for the total stellar mass of the nuclear disc and the underlying main population are reliable. In addition, this also shows that the methodology described above to disentangle the light of the nuclear and underlying discs is trustworthy. Interestingly, our mass estimates indicate that the nuclear disc dominates the stellar mass budget in the central region, with the nuclear disc being $\sim$40\% more massive than the underlying main disc in the same region.

\begin{figure*}
\centering
\includegraphics[width=0.8\linewidth]{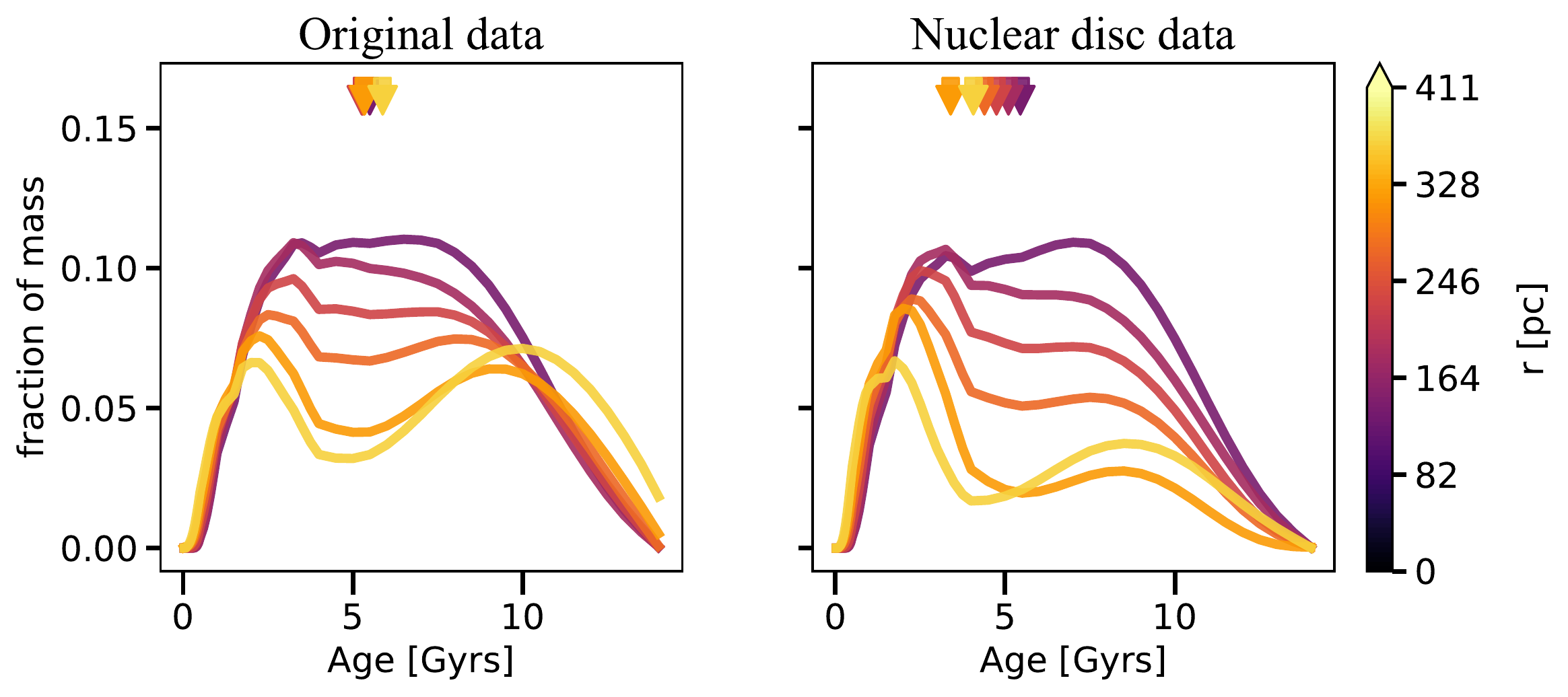}
    \caption{{Star formation histories for NGC\,1433 in different galactocentric radii for the original data (\textbf{left}) and the nuclear disc data (\textbf{right}). In the y-axis we display the fraction of mass formed. On the upper part of each panel we use arrows to display the mean age for each SFH, following the same color coding with respect to the distance to the centre}. For both data cubes, it is clear that the further from the centre, the younger the SFH is (with the exception of the last curve), meaning that the fraction of younger stars increases. This is shown even more strikingly with the subtracted data cube: it is clear that the outskirts of the nuclear disc are in fact younger, in accordance with the inside-out growth scenario of \citet{bittner2019gist}.}
    \label{fig_SFHradial}
\end{figure*}

In addition, we analyse the SFH in different radial bins inside the nuclear disc region for the original data (left panel) and the nuclear disc cleaned data (right panel; Fig. \ref{fig_SFHradial}). At the top of each panel we display the mean age for each radius, colour-coded according to distance from the centre. As one can see for the cleaned ND results, the SFH and the mean ages gradually get younger at larger radii, in agreement with the inside-out growth scenario (\citealp{bittner2020inside}). We discuss the implications of these findings further in Section \ref{subsecInsideOut}.

% %-------------------------------------------------------------------

\section{Discussion}
\label{secDisc}

In this Section, we discuss the results from applying our methodology to the galaxy NGC\,1433 and what they reveal about the formation and evolution of bars and nuclear discs.

\subsection{The old bar in NGC\,1433}
\label{sec:oldbar}

Applying the methodology described in Sect.~\ref{sec:DisLight} and considering our chosen criterion, we find a bar age of $8.0^{+1.6}_{-1.1}\rm{(sys)}^{+0.2}_{-0.5}\rm{(stat)}$ Gyr for NGC\,1433, which hosts a nuclear disc with a radius of 440 pc (e.g., \citealp{gadotti2019time}; \citealp{bittner2020inside}). This corresponds to a redshift $z\approx1$ and is consistent with other observational studies, such as \cite{simmons2014galaxy}, who also find massive galaxies with bars at $z\approx 2$. In addition, \cite{gadotti2015muse} and \cite{perez2017observational} also inferred similar bar ages for other local galaxies, using independent methodologies. Furthermore, this is comparable to predictions from cosmological simulations, which also find bars that form between redshifts $1-2$ and that survive down to $z=0$ (e.g. \citealt{Kraljicetal2012,Fragkoudietal2020}). Our results indicate that NGC\,1433 hosts an old and long-lived bar, which is in accordance with several observed characteristics of this galaxy and our current understanding of bar formation and bar-driven evolutionary processes.

As the bar evolves, the shape of its light profile may change. \cite{kim2015mass} suggested that the bar S\'ersic index ($n_{\mathrm{bar}}$) evolves from an exponential ($n_{\mathrm{bar}} \approx 1-2$) to a flat ($n_{\mathrm{bar}} \approx 0.2$) profile (but see also \citealp{anderson2022secular}). They also measured $n_{\mathrm{bar}}$ for 144 local galaxies from the S$^4$G sample and found that the oldest bars have a $n_{\mathrm{bar}}$ index close to $0.2$. In addition, simulations demonstrate that, as the bar grows older, it becomes more elongated and stronger (e.g., \citealp{athanassoula2013bar}), which can be measured by the bar length $\mathrm{R}_{\mathrm{bar}}$ and the $A_2$ index. The latter is the second component of the Fourier light/mass distribution in the galaxy and is associated with the bar non-axisymmetry. Lastly, the bar-to-total (Bar/T) luminosity ratio also increases as the bar grows longer and more massive, and the bar morphology evolves to a more rectangular or boxy shape (e.g., \citealp{kim2015mass}). The bar morphology is measured by the index $c$, where $c=2.0$ represents a perfect ellipse and $c\geq2.7$ a strongly boxy bar \citep[see][]{gadotti2011}. In summary, an old bar is expected to have a flat light profile ($n_{\mathrm{bar}} \leq 0.7$), high strength ($A_2 \geq 0.4$), a strongly boxy shape ($c \geq 2.7$), relatively large bar-to-total ratio (Bar/T $\geq 0.09$) and bar length normalised by the disc scale-length $\mathrm{R}_{\mathrm{bar}}/h \geq 1.5$. Apart from $A_2$, these values are the median values of the corresponding distributions as found in \citet{gadotti2011} for a sample of about 300 massive barred galaxies.

NGC\,1433 has $n_{\mathrm{bar}}=0.3$, $A_2 = 0.56$, a deprojected, normalised bar length of $1.2$, a boxy shape with $c = 2.9$, and lastly, a Bar/T $ = 0.08$ (see \citealp{kim2014unveiling} and \citealp{diaz-garcia2016}). Most of these characteristics are consistent with the expectation of an old bar, also in accordance with our bar age estimate of $8.0^{+1.6}_{-1.1}\rm{(sys)}^{+0.2}_{-0.5}\rm{(stat)}$ Gyr. Lastly, NGC\,1433 has similar characteristics as NGC\,4371, that was studied by \cite{gadotti2015muse} by also using archaeological evidence present on its nuclear disc. The authors investigated the oldest stars in the nuclear disc to estimate a bar age of $\approx10$ Gyr. In addition, NGC\,4371 has $n_{\mathrm{bar}} = 0.2$, $A_2 = 0.62$, deprojected normalised bar length of $1.3$, $c = 2.7$, and Bar/T $ = 0.08$ (see \citealp{gadotti2015muse} and \citealp{diaz-garcia2016}). Both the age and the characteristics of NGC\,1433 and NGC\,4371 are very consistent with each other and with the scenario whereby they are amongst the first galaxies to form bars. 

The conditions to form a bar are connected with the main disc properties, and depend on the galaxy assembling enough mass and settling in more ordered motion. With this in mind, one can expect that the more massive galaxies will reach the necessary conditions first, following a downsizing picture (e.g., \citealp{sheth2012hot}). Despite that, another plausible possibility is the one where early interactions and/or minor mergers, that happened after the disc has settled, could have triggered the bar formation (e.g., \citealp{noguchi1987close}; \citealp{gerin1989influence}; \citealp{miwa1998dynamical}; \citealp{peschken2019tidally}; \citealp{lokas2021bar}). \cite{gadotti2015muse} argued how this could be possible for NGC\,4371, which is member of the Virgo cluster. Similarly, NGC\,1433 is part of the Dorado group (\citealp{maia1989catalog}). Following the pre-processing picture, galaxies enter clusters with already some level of processing due to earlier interactions (\citealp{haines2015locuss}) while in groups, which could also be responsible for early bar formation $\sim 8$ Gyrs ago. \citet{mendez2010galaxies} and \citet{mendez2012nature} found that galaxies in the Coma and Virgo clusters that host bars are mainly massive ($10^{9} \leq$ M/M$_\odot$ $\leq 10^{11}$). 

\subsection{The inside-out scenario for the growth of nuclear discs}
\label{subsecInsideOut}

\cite{bittner2020inside} showed that in the TIMER sample, the derived age profiles of nuclear discs follow a negative gradient, with the outer parts of the nuclear disc being younger. This is in accordance with an inside-out growth scenario, in which one can have a growing gaseous nuclear disc that forms stars, and/or star formation concentrated in rings of increasing size, on the external borders of the nuclear disc, as shown by H$\alpha$ maps (\citealp{bittner2020inside}). In this scenario, the gas funneled by the bar towards the centre piles up near the inner Lindblad resonance (ILR) of the bar. As the bar grows longer, the radius of the ILR increases, which leads to the build-up of a nuclear disc in an inside-out fashion. 

In order to test this scenario, we analyse the SFHs of the subtracted nuclear disc in different radial bins, as shown in Fig. ~\ref{fig_SFHradial}, and derive the mean ages at each radius (shown as arrows on the top part of the panels). For the original data, the inside-out evidence is subtle, with little change in the mean ages in each radial bin. Nonetheless, with the subtracted data, the inside-out growth of the disc becomes strikingly evident, since in the outer part of the nuclear disc the older stellar populations are almost completely absent from the clean nuclear disc. This is reflected in the mean age at different radii for the subtracted nuclear disc, in which there is a clear gradient towards younger ages at larger radii. This result is also testament that the methodology we develop to subtract the underlying disc component is robust. It shows that the nuclear disc of NGC\,1433 is consistent with the inside-out growth picture for nuclear discs. This highlights how inner structures such as nuclear discs might be assembled in a self-similar way to the corresponding larger scale structure of the main disc. 

Finally, we note that Fig \ref{fig_SFHradial} shows a prominent peak in the SFHs at all radii at young ages ($\sim2.5$Gyr), which is also evident in Fig. \ref{fig_SFH} -- i.e. there is a late burst of star formation, which occurs at younger ages than the first burst associated to the formation of the nuclear disc at $\sim8$Gyr. This implies an event which leads to a renewed inflow of gas at late times, which gives rise to such a burst of star formation. Various mechanisms could give rise to such a late gas inflow event through the bar, such as mechanisms which remove angular momentum from gas, e.g. an interaction or fly-by, or --interestingly for this galaxy -- this late inflow could be related to the buckling of the bar and the formation of the boxy/peanut bulge (e.g. \citealp{perez2017observational}).

\section{Summary and concluding remarks}
\label{sec:summary}

In this study, we develop and present a new method for dating the bar formation epoch of observed disc galaxies. We summarize this work as follows:

\begin{itemize}
    \item We present a new methodology that allows us to disentangle the light from nuclear discs, which are formed by the bar, from the underlying main discs of galaxies, using high-resolution integral field spectroscopic data from MUSE on the VLT. This allows us to find the time at which the nuclear disc formed, by isolating the moment when its star formation starts to dominate over star formation in the underlying population. As nuclear discs are formed due to bar-driven inflow -- which is concurrent with the formation time of the bar -- this allows us to determine the age of the bar. 
    
    \item We perform a number of tests on our methodology, both on observed data, as well as on a hydrodynamic simulation of a barred galaxy (which self-consistently forms a nuclear disc), in order to validate the robustness of the methodology. 
    
    \item As a pilot study, we apply our methodology to the barred galaxy NGC\,1433 from the TIMER survey \citep{gadotti2019time}, and find a bar age of $8.0^{+1.6}_{-1.1}\rm{(sys)}^{+0.2}_{-0.5}\rm{(stat)}$ Gyr. This implies that NGC\,1433 has an old bar, which formed around $z \sim 1$. This aligns with a number of observational characteristics of the galaxy, such as its mass, bar strength, light profile etc. (see Section~\ref{sec:oldbar} for a more detailed discussion on this). Our results are consistent with other studies in the literature which find old bars (e.g. \citealt{simmons2014galaxy,gadotti2015muse}), as well as results from cosmological simulations (e.g. \citealt{Kraljicetal2012,Fragkoudietal2020}), implying that bars can be old, long-lived structures.
    
    \item By examining the SFH of the nuclear disc of NGC\,1433 at different radii, we find that the disc grows inside-out, with younger stars forming at progressively larger radii, with the youngest stars forming at the edge of the nuclear disc, i.e., at the location of the nuclear ring. This is in agreement with an inside-out growth scenario for nuclear discs (see \citealt{bittner2020inside}, and Section~\ref{subsecInsideOut}).
\end{itemize}

We will apply the methodology presented here to the full TIMER sample of barred galaxies, which will provide, for the first time, robust age determinations for bars in a sizable sample of disc galaxies. This will enable us to compare the age of bars with various galaxy properties, such as the total mass, the bar length and pattern speed, enabling us to place important constraints on the evolution of dynamical properties of disc galaxies with time, the epoch of disc settling and the effects of bar-driven evolution.

% %-------------------------------------------------------------------

\begin{acknowledgements}
Based on observations collected at the European Southern Observatory under ESO programmes
296.B-5054, 097.B-0640 and 099.B-0242. Raw and reduced data are available at the ESO Science Archive Facility. AdLC acknowledges financial support from the Spanish Ministry of Science and Innovation (MICINN) through the Spanish State Research Agency, under Severo Ochoa Centres of Excellence Programme 2020-2023 (CEX2019-000920-S). KF acknowledges support through the ESA research fellowship program. PC acknowledges support from Fundação de Amparo à Pesquisa do Estado de São Paulo (FAPESP, 2021/08813-7) and Conselho Nacional de Desenvolvimento Científico e Tecnológico (CNPq, 310555/2021-3). GG acknowledges support from Coordenação de Aperfeiçoamento de Pessoal de Nível Superior (CAPES). TK was supported by the Basic Science Research Program through the National Research Foundation of Korea (NRF) grants (No. 2019R1I1A3A02062242) funded by the Ministry of Education, grants (No. 2022R1A4A3031306) funded by the Korean government (MSIT), and grants (WISET 2021-541) funded by the Korea Foundation for Women In Science, Engineering and Technology. J.M.A. acknowledges the support of the Viera y Clavijo Senior program funded by ACIISI and ULL. PSB acknowledges financial support from the Spanish Ministry of Science, Innovation and Universities (MCIUN) under grant number PID2019-107427GB-C31. GvdV acknowledges funding from the European Research Council (ERC) under the European Union's Horizon 2020 research and innovation programme under grant agreement No 724857 (Consolidator Grant ArcheoDyn).

\end{acknowledgements}

%-------------------------------------------------------------------

\bibliographystyle{aa} % style aa.bst
\bibliography{bibliography} % your references Yourfile.bib

\begin{thebibliography}{105}
\expandafter\ifx\csname natexlab\endcsname\relax\def\natexlab#1{#1}\fi

\bibitem[{Aguerri {et~al.}(2009)Aguerri, M{\'e}ndez-Abreu, \&
  Corsini}]{aguerri2009population}
Aguerri, J., M{\'e}ndez-Abreu, J., \& Corsini, E. 2009, Astronomy \&
  Astrophysics, 495, 491

\bibitem[{Anderson {et~al.}(2022)Anderson, Debattista, Erwin, Liddicott, Deg,
  \& Beraldo~e Silva}]{anderson2022secular}
Anderson, S.~R., Debattista, V.~P., Erwin, P., {et~al.} 2022, Monthly Notices
  of the Royal Astronomical Society, 513, 1642

\bibitem[{{Athanassoula}(1992{\natexlab{a}})}]{athanassoula1992morphology}
{Athanassoula}, E. 1992{\natexlab{a}}, \mnras, 259, 328

\bibitem[{{Athanassoula}(1992{\natexlab{b}})}]{Athanassoula1992b}
{Athanassoula}, E. 1992{\natexlab{b}}, \mnras, 259, 345

\bibitem[{Athanassoula {et~al.}(2013)Athanassoula, Machado, \&
  Rodionov}]{athanassoula2013bar}
Athanassoula, E., Machado, R.~E., \& Rodionov, S. 2013, Monthly Notices of the
  Royal Astronomical Society, 429, 1949

\bibitem[{Athanassoula(2003)}]{athanassoula2003angular}
Athanassoula, L. 2003, in Galaxies and Chaos (Springer), 313--326

\bibitem[{Baba \& Kawata(2020)}]{baba2020age}
Baba, J. \& Kawata, D. 2020, Monthly Notices of the Royal Astronomical Society,
  492, 4500

\bibitem[{Baldwin {et~al.}(1981)Baldwin, Phillips, \&
  Terlevich}]{baldwin1981classification}
Baldwin, J.~A., Phillips, M.~M., \& Terlevich, R. 1981, Publications of the
  Astronomical Society of the Pacific, 93, 5

\bibitem[{Barazza {et~al.}(2008)Barazza, Jogee, \& Marinova}]{barazza2008bars}
Barazza, F.~D., Jogee, S., \& Marinova, I. 2008, The Astrophysical Journal,
  675, 1194

\bibitem[{Bittner {et~al.}(2019)Bittner, Falc{\'o}n-Barroso, Nedelchev, Dorta,
  Gadotti, Sarzi, Molaeinezhad, Iodice, Rosado-Belza, de~Lorenzo-C{\'a}ceres,
  {et~al.}}]{bittner2019gist}
Bittner, A., Falc{\'o}n-Barroso, J., Nedelchev, B., {et~al.} 2019, Astronomy \&
  Astrophysics, 628, A117

\bibitem[{Bittner {et~al.}(2020)Bittner, S{\'a}nchez-Bl{\'a}zquez, Gadotti,
  Neumann, Fragkoudi, Coelho, de~Lorenzo-C{\'a}ceres, Falc{\'o}n-Barroso, Kim,
  Leaman, {et~al.}}]{bittner2020inside}
Bittner, A., S{\'a}nchez-Bl{\'a}zquez, P., Gadotti, D.~A., {et~al.} 2020,
  Astronomy \& Astrophysics, 643, A65

\bibitem[{Breda {et~al.}(2020)Breda, Papaderos, \&
  Gomes}]{breda2020indications}
Breda, I., Papaderos, P., \& Gomes, J.-M. 2020, Astronomy \& Astrophysics, 640,
  A20

\bibitem[{Buta {et~al.}(2015)Buta, Sheth, Athanassoula, Bosma, Knapen,
  Laurikainen, Salo, Elmegreen, Ho, Zaritsky, {et~al.}}]{buta2015classical}
Buta, R.~J., Sheth, K., Athanassoula, E., {et~al.} 2015, The Astrophysical
  Journal Supplement Series, 217, 32

\bibitem[{Cappellari(2012)}]{cappellari2012ppxf}
Cappellari, M. 2012, Astrophysics Source Code Library, ascl

\bibitem[{Cappellari(2017)}]{cappellari2017improving}
Cappellari, M. 2017, Monthly Notices of the Royal Astronomical Society, 466,
  798

\bibitem[{Cappellari \& Copin(2003)}]{cappellari2003adaptive}
Cappellari, M. \& Copin, Y. 2003, Monthly Notices of the Royal Astronomical
  Society, 342, 345

\bibitem[{Cappellari \& Emsellem(2004)}]{cappellari2004parametric}
Cappellari, M. \& Emsellem, E. 2004, Publications of the Astronomical Society
  of the Pacific, 116, 138

\bibitem[{Coelho \& Gadotti(2011)}]{coelho2011bars}
Coelho, P. \& Gadotti, D.~A. 2011, The Astrophysical Journal Letters, 743, L13

\bibitem[{Combes \& Gerin(1985)}]{combes1985spiral}
Combes, F. \& Gerin, M. 1985, Astronomy and Astrophysics, 150, 327

\bibitem[{Cresci {et~al.}(2009)Cresci, Hicks, Genzel, Schreiber, Davies,
  Bouch{\'e}, Buschkamp, Genel, Shapiro, Tacconi, {et~al.}}]{cresci2009sins}
Cresci, G., Hicks, E., Genzel, R., {et~al.} 2009, The Astrophysical Journal,
  697, 115

\bibitem[{de~Lorenzo-C{\'a}ceres {et~al.}(2019)de~Lorenzo-C{\'a}ceres,
  S{\'a}nchez-Bl{\'a}zquez, M{\'e}ndez-Abreu, Gadotti, Falc{\'o}n-Barroso,
  Mart{\'\i}nez-Valpuesta, Coelho, Fragkoudi, Husemann, Leaman,
  {et~al.}}]{de2019clocking}
de~Lorenzo-C{\'a}ceres, A., S{\'a}nchez-Bl{\'a}zquez, P., M{\'e}ndez-Abreu, J.,
  {et~al.} 2019, Monthly Notices of the Royal Astronomical Society, 484, 5296

\bibitem[{Dekel {et~al.}(2009)Dekel, Ceverino, {et~al.}}]{dekel2009formation}
Dekel, A., Ceverino, D., {et~al.} 2009, The Astrophysical Journal, 703, 785

\bibitem[{{Di Matteo} {et~al.}(2013){Di Matteo}, {Haywood}, {Combes},
  {Semelin}, \& {Snaith}}]{dimatteoetal2013}
{Di Matteo}, P., {Haywood}, M., {Combes}, F., {Semelin}, B., \& {Snaith}, O.~N.
  2013, \aap, 553, A102

\bibitem[{{D{\'\i}az-Garc{\'\i}a} {et~al.}(2016){D{\'\i}az-Garc{\'\i}a},
  {Salo}, {Laurikainen}, \& {Herrera-Endoqui}}]{diaz-garcia2016}
{D{\'\i}az-Garc{\'\i}a}, S., {Salo}, H., {Laurikainen}, E., \&
  {Herrera-Endoqui}, M. 2016, \aap, 587, A160

\bibitem[{Ellison {et~al.}(2011)Ellison, Nair, Patton, Scudder, Mendel, \&
  Simard}]{ellison2011impact}
Ellison, S.~L., Nair, P., Patton, D.~R., {et~al.} 2011, Monthly Notices of the
  Royal Astronomical Society, 416, 2182

\bibitem[{Elmegreen \& Elmegreen(2006)}]{elmegreen2006observations}
Elmegreen, B.~G. \& Elmegreen, D.~M. 2006, The Astrophysical Journal, 650, 644

\bibitem[{Emsellem {et~al.}(2015)Emsellem, Renaud, Bournaud, Elmegreen, Combes,
  \& Gabor}]{emsellem2015interplay}
Emsellem, E., Renaud, F., Bournaud, F., {et~al.} 2015, Monthly Notices of the
  Royal Astronomical Society, 446, 2468

\bibitem[{Emsellem {et~al.}(2021)Emsellem, Schinnerer, Santoro, Belfiore,
  Pessa, McElroy, Blanc, Congiu, Groves, Ho, {et~al.}}]{emsellem2021phangs}
Emsellem, E., Schinnerer, E., Santoro, F., {et~al.} 2021, arXiv preprint
  arXiv:2110.03708

\bibitem[{Epinat {et~al.}(2012)Epinat, Tasca, Amram, Contini, Le~F{\`e}vre,
  Queyrel, Vergani, Garilli, Kissler-Patig, Moultaka,
  {et~al.}}]{epinat2012massiv}
Epinat, B., Tasca, L., Amram, P., {et~al.} 2012, Astronomy \& Astrophysics,
  539, A92

\bibitem[{Erwin(2018)}]{erwin2018dependence}
Erwin, P. 2018, Monthly Notices of the Royal Astronomical Society, 474, 5372

\bibitem[{Eskridge {et~al.}(2000)Eskridge, Frogel, Pogge, Quillen, Davies,
  DePoy, Houdashelt, Kuchinski, Ram{\'\i}rez, Sellgren,
  {et~al.}}]{eskridge2000frequency}
Eskridge, P.~B., Frogel, J.~A., Pogge, R.~W., {et~al.} 2000, The Astronomical
  Journal, 119, 536

\bibitem[{Falc{\'o}n-Barroso {et~al.}(2002)Falc{\'o}n-Barroso, Peletier, \&
  Balcells}]{falcon2002bulges}
Falc{\'o}n-Barroso, J., Peletier, R.~F., \& Balcells, M. 2002, Monthly Notices
  of the Royal Astronomical Society, 335, 741

\bibitem[{{Ferreira} {et~al.}(2022){Ferreira}, {Adams}, {Conselice},
  {Sazonova}, {Austin}, {Caruana}, {Ferrari}, {Broadhurst}, {Diego}, {Frye},
  {Pascale}, {Wilkins}, {Windhorst}, \& {Zitrin}}]{2022arXiv220709428F}
{Ferreira}, L., {Adams}, N., {Conselice}, C.~J., {et~al.} 2022, arXiv e-prints,
  arXiv:2207.09428

\bibitem[{{Fragkoudi} {et~al.}(2016){Fragkoudi}, {Athanassoula}, \&
  {Bosma}}]{fragkoudietal2016}
{Fragkoudi}, F., {Athanassoula}, E., \& {Bosma}, A. 2016, \mnras, 462, L41

\bibitem[{{Fragkoudi} {et~al.}(2017){Fragkoudi}, {Di Matteo}, {Haywood},
  {G{\'o}mez}, {Combes}, {Katz}, \& {Semelin}}]{fragkoudietal2017b}
{Fragkoudi}, F., {Di Matteo}, P., {Haywood}, M., {et~al.} 2017, \aap, 606, A47

\bibitem[{{Fragkoudi} {et~al.}(2020){Fragkoudi}, {Grand}, {Pakmor},
  {Bl{\'a}zquez-Calero}, {Gargiulo}, {Gomez}, {Marinacci}, {Monachesi}, {Ness},
  {Perez}, {Tissera}, \& {White}}]{Fragkoudietal2020}
{Fragkoudi}, F., {Grand}, R.~J.~J., {Pakmor}, R., {et~al.} 2020, \mnras, 494,
  5936

\bibitem[{{Fragkoudi} {et~al.}(2021){Fragkoudi}, {Grand}, {Pakmor}, {Springel},
  {White}, {Marinacci}, {Gomez}, \& {Navarro}}]{fragkoudietal2021}
{Fragkoudi}, F., {Grand}, R.~J.~J., {Pakmor}, R., {et~al.} 2021, \aap, 650, L16

\bibitem[{{Gadotti}(2011)}]{gadotti2011}
{Gadotti}, D.~A. 2011, \mnras, 415, 3308

\bibitem[{Gadotti {et~al.}(2020)Gadotti, Bittner, Falc{\'o}n-Barroso,
  M{\'e}ndez-Abreu, Kim, Fragkoudi, de~Lorenzo-C{\'a}ceres, Leaman, Neumann,
  Querejeta, {et~al.}}]{gadotti2020kinematic}
Gadotti, D.~A., Bittner, A., Falc{\'o}n-Barroso, J., {et~al.} 2020, Astronomy
  \& Astrophysics, 643, A14

\bibitem[{Gadotti {et~al.}(2019)Gadotti, S{\'a}nchez-Bl{\'a}zquez,
  Falc{\'o}n-Barroso, Husemann, Seidel, P{\'e}rez, de~Lorenzo-C{\'a}ceres,
  Martinez-Valpuesta, Fragkoudi, Leung, {et~al.}}]{gadotti2019time}
Gadotti, D.~A., S{\'a}nchez-Bl{\'a}zquez, P., Falc{\'o}n-Barroso, J., {et~al.}
  2019, Monthly Notices of the Royal Astronomical Society, 482, 506

\bibitem[{Gadotti {et~al.}(2015)Gadotti, Seidel, S{\'a}nchez-Bl{\'a}zquez,
  Falc{\'o}n-Barroso, Husemann, Coelho, \& P{\'e}rez}]{gadotti2015muse}
Gadotti, D.~A., Seidel, M.~K., S{\'a}nchez-Bl{\'a}zquez, P., {et~al.} 2015,
  Astronomy \& Astrophysics, 584, A90

\bibitem[{Gao {et~al.}(2019)Gao, Ho, Barth, \& Li}]{gao2019carnegie}
Gao, H., Ho, L.~C., Barth, A.~J., \& Li, Z.-Y. 2019, The Astrophysical Journal
  Supplement Series, 244, 34

\bibitem[{Genzel {et~al.}(2008)Genzel, Burkert, Bouch{\'e}, Cresci, Schreiber,
  Shapley, Shapiro, Tacconi, Buschkamp, Cimatti, {et~al.}}]{genzel2008rings}
Genzel, R., Burkert, A., Bouch{\'e}, N., {et~al.} 2008, The Astrophysical
  Journal, 687, 59

\bibitem[{Gerin {et~al.}(1989)Gerin, Combes, \&
  Athanassoula}]{gerin1989influence}
Gerin, M., Combes, F., \& Athanassoula, E. 1989, Dynamics of Astrophysical
  Discs, 219

\bibitem[{G{\'e}ron {et~al.}(2021)G{\'e}ron, Smethurst, Lintott, Kruk, Masters,
  Simmons, \& Stark}]{geron2021galaxy}
G{\'e}ron, T., Smethurst, R.~J., Lintott, C., {et~al.} 2021, Monthly Notices of
  the Royal Astronomical Society, 507, 4389

\bibitem[{Haines {et~al.}(2015)Haines, Pereira, Smith, Egami, Babul,
  Finoguenov, Ziparo, McGee, Rawle, Okabe, {et~al.}}]{haines2015locuss}
Haines, C., Pereira, M., Smith, G.~P., {et~al.} 2015, The Astrophysical
  Journal, 806, 101

\bibitem[{{Halle} {et~al.}(2015){Halle}, {Di Matteo}, {Haywood}, \&
  {Combes}}]{halleetal2015}
{Halle}, A., {Di Matteo}, P., {Haywood}, M., \& {Combes}, F. 2015, \aap, 578,
  A58

\bibitem[{Haywood {et~al.}(2016)Haywood, Lehnert, Di~Matteo, Snaith,
  Schultheis, Katz, \& G{\'o}mez}]{haywood2016milky}
Haywood, M., Lehnert, M., Di~Matteo, P., {et~al.} 2016, Astronomy \&
  Astrophysics, 589, A66

\bibitem[{Ho {et~al.}(2011)Ho, Li, Barth, Seigar, \& Peng}]{ho2011carnegie}
Ho, L.~C., Li, Z.-Y., Barth, A.~J., Seigar, M.~S., \& Peng, C.~Y. 2011, The
  Astrophysical Journal Supplement Series, 197, 21

\bibitem[{Ishizuki {et~al.}(1990)Ishizuki, Kawabe, Ishiguro, Okumura, Morita,
  Chikada, \& Kasuga}]{ishizuki1990molecular}
Ishizuki, S., Kawabe, R., Ishiguro, M., {et~al.} 1990, Nature, 344, 224

\bibitem[{Kim {et~al.}(2014)Kim, Gadotti, Sheth, Athanassoula, Bosma, Lee,
  Madore, Elmegreen, Knapen, Zaritsky, {et~al.}}]{kim2014unveiling}
Kim, T., Gadotti, D.~A., Sheth, K., {et~al.} 2014, The Astrophysical Journal,
  782, 64

\bibitem[{Kim {et~al.}(2015)Kim, Sheth, Gadotti, Lee, Zaritsky, Elmegreen,
  Athanassoula, Bosma, Holwerda, Ho, {et~al.}}]{kim2015mass}
Kim, T., Sheth, K., Gadotti, D.~A., {et~al.} 2015, The Astrophysical Journal,
  799, 99

\bibitem[{{Kimm} \& {Cen}(2014)}]{KimmCen2014}
{Kimm}, T. \& {Cen}, R. 2014, \apj, 788, 121

\bibitem[{{Kimm} {et~al.}(2015){Kimm}, {Cen}, {Devriendt}, {Dubois}, \&
  {Slyz}}]{Kimmetal2015}
{Kimm}, T., {Cen}, R., {Devriendt}, J., {Dubois}, Y., \& {Slyz}, A. 2015,
  \mnras, 451, 2900

\bibitem[{Kormendy \& Kennicutt~Jr(2004)}]{kormendy2004secular}
Kormendy, J. \& Kennicutt~Jr, R.~C. 2004, arXiv preprint astro-ph/0407343

\bibitem[{{Kraljic} {et~al.}(2012){Kraljic}, {Bournaud}, \&
  {Martig}}]{Kraljicetal2012}
{Kraljic}, K., {Bournaud}, F., \& {Martig}, M. 2012, \apj, 757, 60

\bibitem[{Kraljic {et~al.}(2012)Kraljic, Bournaud, \& Martig}]{kraljic2012two}
Kraljic, K., Bournaud, F., \& Martig, M. 2012, The Astrophysical Journal, 757,
  60

\bibitem[{Kroupa(2001)}]{kroupa2001variation}
Kroupa, P. 2001, Monthly Notices of the Royal Astronomical Society, 322, 231

\bibitem[{Law {et~al.}(2009)Law, Steidel, Erb, Larkin, Pettini, Shapley, \&
  Wright}]{law2009kiloparsec}
Law, D.~R., Steidel, C.~C., Erb, D.~K., {et~al.} 2009, The Astrophysical
  Journal, 697, 2057

\bibitem[{Lelli {et~al.}(2021)Lelli, Di~Teodoro, Fraternali, Man, Zhang,
  De~Breuck, Davis, \& Maiolino}]{lelli2021massive}
Lelli, F., Di~Teodoro, E.~M., Fraternali, F., {et~al.} 2021, Science, 371, 713

\bibitem[{{\L}okas(2021)}]{lokas2021bar}
{\L}okas, E.~L. 2021, Astronomy \& Astrophysics, 647, A143

\bibitem[{Lynden-Bell \& Kalnajs(1972)}]{lynden1972generating}
Lynden-Bell, D. \& Kalnajs, A. 1972, Monthly Notices of the Royal Astronomical
  Society, 157, 1

\bibitem[{Maia {et~al.}(1989)Maia, Da~Costa, \& Latham}]{maia1989catalog}
Maia, M., Da~Costa, L., \& Latham, D.~W. 1989, The Astrophysical Journal
  Supplement Series, 69, 809

\bibitem[{Masters {et~al.}(2012)Masters, Nichol, Haynes, Keel, Lintott,
  Simmons, Skibba, Bamford, Giovanelli, \& Schawinski}]{masters2012galaxy}
Masters, K.~L., Nichol, R.~C., Haynes, M.~P., {et~al.} 2012, Monthly Notices of
  the Royal Astronomical Society, 424, 2180

\bibitem[{McDermid {et~al.}(2015)McDermid, Alatalo, Blitz, Bournaud, Bureau,
  Cappellari, Crocker, Davies, Davis, De~Zeeuw, {et~al.}}]{mcdermid2015atlas}
McDermid, R.~M., Alatalo, K., Blitz, L., {et~al.} 2015, Monthly Notices of the
  Royal Astronomical Society, 448, 3484

\bibitem[{M{\'e}ndez-Abreu {et~al.}(2010)M{\'e}ndez-Abreu, S{\'a}nchez-Janssen,
  \& Aguerri}]{mendez2010galaxies}
M{\'e}ndez-Abreu, J., S{\'a}nchez-Janssen, R., \& Aguerri, J. 2010, The
  Astrophysical Journal Letters, 711, L61

\bibitem[{M{\'e}ndez-Abreu {et~al.}(2012)M{\'e}ndez-Abreu, S{\'a}nchez-Janssen,
  Aguerri, Corsini, \& Zarattini}]{mendez2012nature}
M{\'e}ndez-Abreu, J., S{\'a}nchez-Janssen, R., Aguerri, J., Corsini, E., \&
  Zarattini, S. 2012, The Astrophysical Journal Letters, 761, L6

\bibitem[{Men{\'e}ndez-Delmestre {et~al.}(2007)Men{\'e}ndez-Delmestre, Sheth,
  Schinnerer, Jarrett, \& Scoville}]{menendez2007near}
Men{\'e}ndez-Delmestre, K., Sheth, K., Schinnerer, E., Jarrett, T.~H., \&
  Scoville, N.~Z. 2007, The Astrophysical Journal, 657, 790

\bibitem[{Miwa \& Noguchi(1998)}]{miwa1998dynamical}
Miwa, T. \& Noguchi, M. 1998, The Astrophysical Journal, 499, 149

\bibitem[{Mu{\~n}oz-Mateos {et~al.}(2015)Mu{\~n}oz-Mateos, Sheth, Regan, Kim,
  Laine, Erroz-Ferrer, De~Paz, Comeron, Hinz, Laurikainen,
  {et~al.}}]{munoz2015spitzer}
Mu{\~n}oz-Mateos, J.~C., Sheth, K., Regan, M., {et~al.} 2015, The Astrophysical
  Journal Supplement Series, 219, 3

\bibitem[{Munoz-Tun{\'o}n {et~al.}(2004)Munoz-Tun{\'o}n, Caon, \&
  Aguerri}]{munoz2004inner}
Munoz-Tun{\'o}n, C., Caon, N., \& Aguerri, J. A.~L. 2004, The Astronomical
  Journal, 127, 58

\bibitem[{Nair \& Abraham(2010)}]{nair2010fraction}
Nair, P.~B. \& Abraham, R.~G. 2010, The Astrophysical Journal Letters, 714,
  L260

\bibitem[{{Navarro} {et~al.}(1997){Navarro}, {Frenk}, \& {White}}]{NFW1997}
{Navarro}, J.~F., {Frenk}, C.~S., \& {White}, S. D.~M. 1997, \apj, 490, 493

\bibitem[{Newman {et~al.}(2013)Newman, Genzel, Schreiber, Griffin, Mancini,
  Lilly, Renzini, Bouch{\'e}, Burkert, Buschkamp, {et~al.}}]{newman2013sins}
Newman, S.~F., Genzel, R., Schreiber, N. M.~F., {et~al.} 2013, The
  Astrophysical Journal, 767, 104

\bibitem[{Noguchi(1987)}]{noguchi1987close}
Noguchi, M. 1987, Monthly Notices of the Royal Astronomical Society, 228, 635

\bibitem[{Oser {et~al.}(2010)Oser, Ostriker, Naab, Johansson, \&
  Burkert}]{oser2010two}
Oser, L., Ostriker, J.~P., Naab, T., Johansson, P.~H., \& Burkert, A. 2010, The
  Astrophysical Journal, 725, 2312

\bibitem[{Papaderos {et~al.}(2021)Papaderos, Breda, Humphrey, Gomes, Ziegler,
  \& Pappalardo}]{papaderos2021inside}
Papaderos, P., Breda, I., Humphrey, A., {et~al.} 2021, arXiv preprint
  arXiv:2111.05200

\bibitem[{P{\'e}rez {et~al.}(2017)P{\'e}rez, Mart{\'\i}nez-Valpuesta,
  Ruiz-Lara, de~Lorenzo-Caceres, Falc{\'o}n-Barroso, Florido,
  Gonz{\'a}lez~Delgado, Lyubenova, Marino, S{\'a}nchez,
  {et~al.}}]{perez2017observational}
P{\'e}rez, I., Mart{\'\i}nez-Valpuesta, I., Ruiz-Lara, T., {et~al.} 2017,
  Monthly Notices of the Royal Astronomical Society: Letters, 470, L122

\bibitem[{{Perret}(2016)}]{Perret2016}
{Perret}, V. 2016, {DICE: Disk Initial Conditions Environment}, Astrophysics
  Source Code Library, record ascl:1607.002

\bibitem[{{Perret} {et~al.}(2014){Perret}, {Renaud}, {Epinat}, {Amram},
  {Bournaud}, {Contini}, {Teyssier}, \& {Lambert}}]{Perretetal2014}
{Perret}, V., {Renaud}, F., {Epinat}, B., {et~al.} 2014, \aap, 562, A1

\bibitem[{Peschken \& {\L}okas(2019)}]{peschken2019tidally}
Peschken, N. \& {\L}okas, E.~L. 2019, Monthly Notices of the Royal Astronomical
  Society, 483, 2721

\bibitem[{Pietrinferni {et~al.}(2004)Pietrinferni, Cassisi, Salaris, \&
  Castelli}]{pietrinferni2004large}
Pietrinferni, A., Cassisi, S., Salaris, M., \& Castelli, F. 2004, The
  Astrophysical Journal, 612, 168

\bibitem[{Pietrinferni {et~al.}(2006)Pietrinferni, Cassisi, Salaris, \&
  Castelli}]{pietrinferni2006large}
Pietrinferni, A., Cassisi, S., Salaris, M., \& Castelli, F. 2006, The
  Astrophysical Journal, 642, 797

\bibitem[{Pietrinferni {et~al.}(2013)Pietrinferni, Cassisi, Salaris, \&
  Hidalgo}]{pietrinferni2013basti}
Pietrinferni, A., Cassisi, S., Salaris, M., \& Hidalgo, S. 2013, Astronomy \&
  Astrophysics, 558, A46

\bibitem[{Pietrinferni {et~al.}(2009)Pietrinferni, Cassisi, Salaris, Percival,
  \& Ferguson}]{pietrinferni2009large}
Pietrinferni, A., Cassisi, S., Salaris, M., Percival, S., \& Ferguson, J.~W.
  2009, The Astrophysical Journal, 697, 275

\bibitem[{Rizzo {et~al.}(2020)Rizzo, Vegetti, Powell, Fraternali, McKean,
  Stacey, \& White}]{rizzo2020dynamically}
Rizzo, F., Vegetti, S., Powell, D., {et~al.} 2020, Nature, 584, 201

\bibitem[{{Romero-G{\'o}mez} {et~al.}(2007){Romero-G{\'o}mez}, {Athanassoula},
  {Masdemont}, \& {Garc{\'\i}a-G{\'o}mez}}]{romerogomezetal2007}
{Romero-G{\'o}mez}, M., {Athanassoula}, E., {Masdemont}, J.~J., \&
  {Garc{\'\i}a-G{\'o}mez}, C. 2007, \aap, 472, 63

\bibitem[{{Rosas-Guevara} {et~al.}(2020){Rosas-Guevara}, {Bonoli}, {Dotti},
  {Zana}, {Nelson}, {Pillepich}, {Ho}, {Izquierdo-Villalba}, {Hernquist}, \&
  {Pakmor}}]{rosasguevaraetal2020}
{Rosas-Guevara}, Y., {Bonoli}, S., {Dotti}, M., {et~al.} 2020, \mnras, 491,
  2547

\bibitem[{Salo {et~al.}(2015)Salo, Laurikainen, Laine, Comer{\'o}n, Gadotti,
  Buta, Sheth, Zaritsky, Ho, Knapen, {et~al.}}]{salo2015spitzer}
Salo, H., Laurikainen, E., Laine, J., {et~al.} 2015, The Astrophysical Journal
  Supplement Series, 219, 4

\bibitem[{Sanchez-Blazquez {et~al.}(2011)Sanchez-Blazquez, Ocvirk, Gibson,
  P{\'e}rez, \& Peletier}]{sanchez2011star}
Sanchez-Blazquez, P., Ocvirk, P., Gibson, B.~K., P{\'e}rez, I., \& Peletier,
  R.~F. 2011, Monthly Notices of the Royal Astronomical Society, 415, 709

\bibitem[{Sarzi {et~al.}(2018)Sarzi, Iodice, Coccato, Corsini, de~Zeeuw,
  Falc{\'o}n-Barroso, Gadotti, Lyubenova, McDermid, van~de Ven,
  {et~al.}}]{sarzi2018fornax3d}
Sarzi, M., Iodice, E., Coccato, L., {et~al.} 2018, Astronomy \& Astrophysics,
  616, A121

\bibitem[{Schawinski {et~al.}(2014)Schawinski, Urry, Simmons, Fortson, Kaviraj,
  Keel, Lintott, Masters, Nichol, Sarzi, {et~al.}}]{schawinski2014green}
Schawinski, K., Urry, C.~M., Simmons, B.~D., {et~al.} 2014, Monthly Notices of
  the Royal Astronomical Society, 440, 889

\bibitem[{Schreiber {et~al.}(2009)Schreiber, Genzel, Bouch{\'e}, Cresci,
  Davies, Buschkamp, Shapiro, Tacconi, Hicks, Genel,
  {et~al.}}]{schreiber2009sins}
Schreiber, N.~F., Genzel, R., Bouch{\'e}, N., {et~al.} 2009, The Astrophysical
  Journal, 706, 1364

\bibitem[{Schreiber {et~al.}(2006)Schreiber, Genzel, Lehnert, Bouch{\'e},
  Verma, Erb, Shapley, Steidel, Davies, Lutz, {et~al.}}]{schreiber2006sinfoni}
Schreiber, N.~F., Genzel, R., Lehnert, M., {et~al.} 2006, The Astrophysical
  Journal, 645, 1062

\bibitem[{Seo {et~al.}(2019)Seo, Kim, Kwak, Hsieh, Han, \&
  Hopkins}]{seo2019effects}
Seo, W.-Y., Kim, W.-T., Kwak, S., {et~al.} 2019, The Astrophysical Journal,
  872, 5

\bibitem[{Shapiro {et~al.}(2008)Shapiro, Genzel, Schreiber, Tacconi,
  Bouch{\'e}, Cresci, Davies, Eisenhauer, Johansson, Krajnovi{\'c},
  {et~al.}}]{shapiro2008kinemetry}
Shapiro, K.~L., Genzel, R., Schreiber, N. M.~F., {et~al.} 2008, The
  Astrophysical Journal, 682, 231

\bibitem[{Sheth {et~al.}(2008)Sheth, Elmegreen, Elmegreen, Capak, Abraham,
  Athanassoula, Ellis, Mobasher, Salvato, Schinnerer,
  {et~al.}}]{sheth2008evolution}
Sheth, K., Elmegreen, D.~M., Elmegreen, B.~G., {et~al.} 2008, The Astrophysical
  Journal, 675, 1141

\bibitem[{Sheth {et~al.}(2012)Sheth, Melbourne, Elmegreen, Elmegreen,
  Athanassoula, Abraham, \& Weiner}]{sheth2012hot}
Sheth, K., Melbourne, J., Elmegreen, D.~M., {et~al.} 2012, The Astrophysical
  Journal, 758, 136

\bibitem[{Sheth {et~al.}(2005)Sheth, Vogel, Regan, Thornley, \&
  Teuben}]{sheth2005secular}
Sheth, K., Vogel, S.~N., Regan, M.~W., Thornley, M.~D., \& Teuben, P.~J. 2005,
  The Astrophysical Journal, 632, 217

\bibitem[{Simmons {et~al.}(2014)Simmons, Melvin, Lintott, Masters, Willett,
  Keel, Smethurst, Cheung, Nichol, Schawinski, {et~al.}}]{simmons2014galaxy}
Simmons, B.~D., Melvin, T., Lintott, C., {et~al.} 2014, Monthly Notices of the
  Royal Astronomical Society, 445, 3466

\bibitem[{{Teyssier}(2002)}]{Teyssier2002}
{Teyssier}, R. 2002, \aap, 385, 337

\bibitem[{Vazdekis {et~al.}(2015)Vazdekis, Coelho, Cassisi, Ricciardelli,
  Falc{\'o}n-Barroso, S{\'a}nchez-Bl{\'a}zquez, Barbera, Beasley, \&
  Pietrinferni}]{vazdekis2015evolutionary}
Vazdekis, A., Coelho, P., Cassisi, S., {et~al.} 2015, Monthly Notices of the
  Royal Astronomical Society, 449, 1177

\bibitem[{Vazdekis {et~al.}(2016)Vazdekis, Koleva, Ricciardelli, R{\"o}ck, \&
  Falc{\'o}n-Barroso}]{vazdekis2016uv}
Vazdekis, A., Koleva, M., Ricciardelli, E., R{\"o}ck, B., \&
  Falc{\'o}n-Barroso, J. 2016, Monthly Notices of the Royal Astronomical
  Society, 463, 3409

\bibitem[{Wisnioski {et~al.}(2015)Wisnioski, Schreiber, Wuyts, Wuyts, Bandara,
  Wilman, Genzel, Bender, Davies, Fossati, {et~al.}}]{wisnioski2015kmos3d}
Wisnioski, E., Schreiber, N.~F., Wuyts, S., {et~al.} 2015, The Astrophysical
  Journal, 799, 209

\bibitem[{Zhu {et~al.}(2018)Zhu, van~de Ven, M{\'e}ndez-Abreu, \&
  Obreja}]{zhu2018morphology}
Zhu, L., van~de Ven, G., M{\'e}ndez-Abreu, J., \& Obreja, A. 2018, Monthly
  Notices of the Royal Astronomical Society, 479, 945

\end{thebibliography}

\begin{appendix}
\label{appendixRef}

\section{Control galaxies -- NGC\,1380 and NGC\,1084}
\label{appdx_control}

In order to assess whether our methodology creates spurious results for galaxies that do not host a nuclear disc, we applied the same methodology for two control galaxies: NGC\,1380 and NGC\,1084. 

Considering NGC\,1380, \cite{gao2019carnegie} describe the galaxy as an inclined system with a classical bulge and no clear presence of a bar. The galaxy is at a distance of 21.2 Mpc with inclination of 47$^\circ$ (see references in \citealp{sarzi2018fornax3d} and \citealp{gao2019carnegie}). We consider the effective radius of the bulge to be 1080 pc (12.1 arcsec) from the photometric decomposition performed in \cite{gao2019carnegie} in lieu of the nuclear disc radius. The data for NGC\,1380 comes from the ESO archive, PI: Sarzi, M., programme ID 296.B-5054, using MUSE in Wide Field Mode. Further details about the galaxy and the observations can be found in \cite{sarzi2018fornax3d}.

As a second control galaxy, NGC\,1084 is a barless galaxy (\citealp{gao2019carnegie}) with inclination\footnote{See http://leda.univ-lyon1.fr/ledacat.cgi} of 49.9$^\circ$ that also does not host a nuclear disc. \cite{gao2019carnegie} also describe its morphology with a broken inner disc and a bulge with effective radius of $\approx$ 150 pc (4.3 arcsec). The data for NGC\,1084 comes from the ESO archive, PI: Carollo, C. M., programme ID 099.B-0242, using MUSE in Wide Field Mode.

\begin{figure}
\centering
\includegraphics[width=0.9\linewidth]{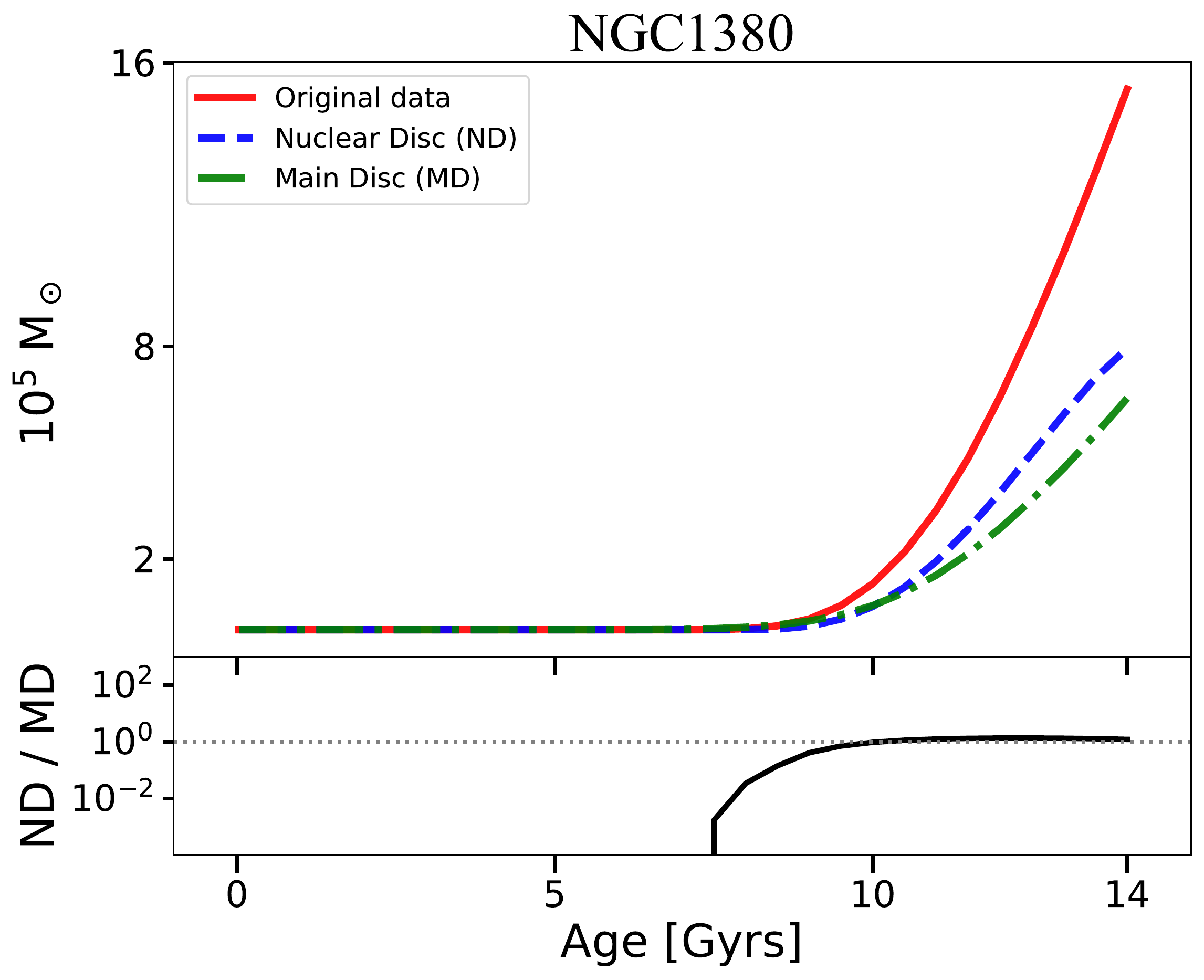}
\includegraphics[width=0.9\linewidth]{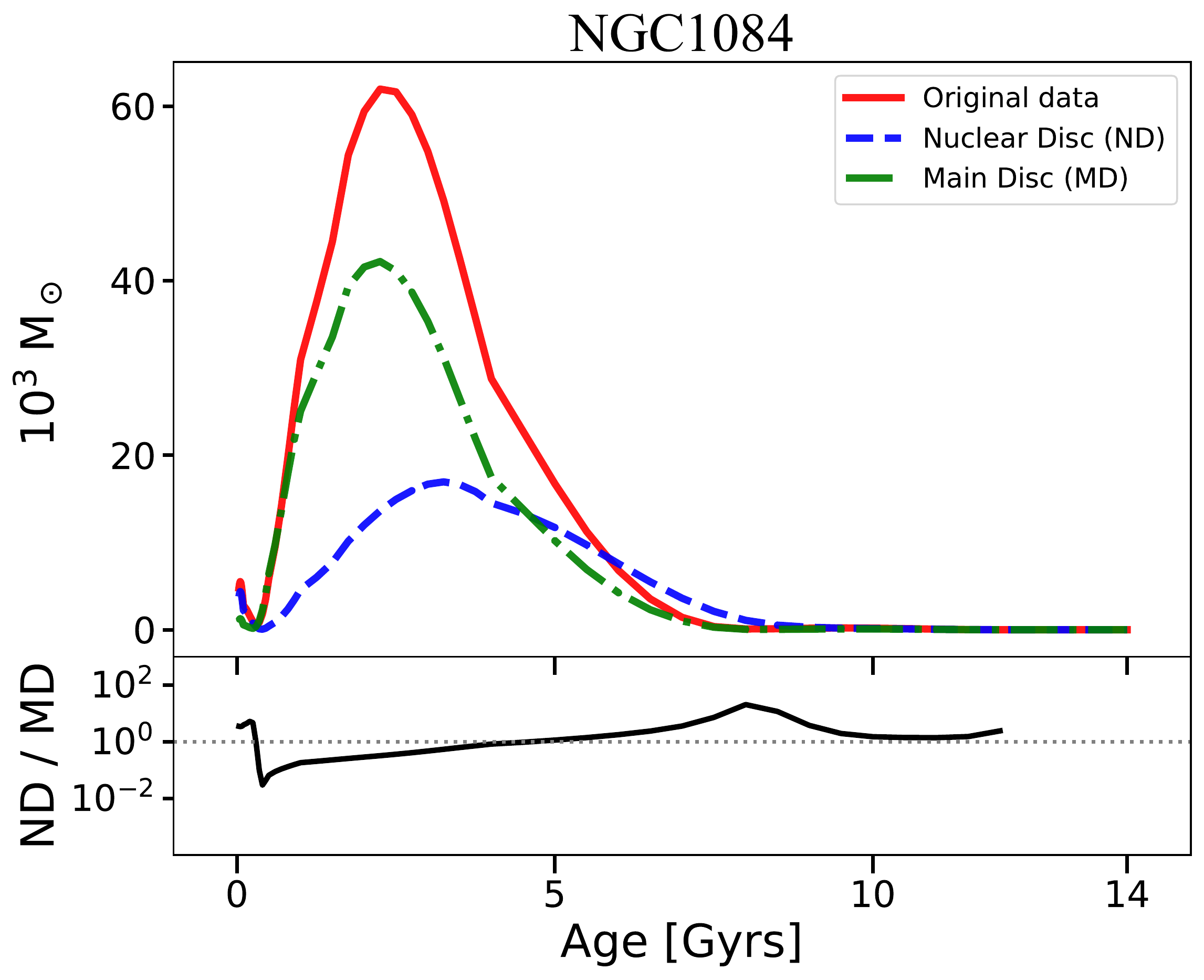}
    \caption{Same as Fig.~\ref{fig_SFH} but for NGC\,1380 (top) and NGC\,1084 (bottom). It is clear that for both galaxies the representative main disc and the "cleaned" data do not show important deviations in their SFHs.}
    \label{fig_NGC1380}
\end{figure}

Figure \ref{fig_NGC1380} shows the outcome of these tests. It is clear that the subtracted spectra show SFHs similar to those in the original spectra. Therefore, our methodology does not artificially produce differences in the SFHs of the regions where the nuclear and main underlying discs dominate, which would wrongly be attributed to the formation of the bar.

% \subsection{NGC\,1084}

%\section{Tests with the simulation}

%\subsection{The age gradient in the disc}

%\subsection{The location of the representative spectrum}

\section{Further tests on the methodology}
\label{app_systematic}

In order to test our methodology and assess the effects of systematic errors, we tested different configurations to assess the possible range of bar ages thus deduced. Our tests include varying the position of the representative ring, using different light profiles to build the underlying main disc, not collapsing the data cubes, but rather deriving a mean SFH from the SFHs corresponding to each individual spaxel in the data cubes, and lastly, changing the \texttt{pPXF} regularisation error parameter.

Considering all the possible different configurations, we estimate the final systematic errors as $8.0^{+1.6}_{-1.1}\rm{(sys)}$ Gyr. In the following we describe each of the tests individually.

\subsection{The location of the representative spectrum}
\label{apdx_ring}

Ideally, the representative ring should be located immediately after the end of the nuclear disc, such that it is not contaminated by light from it, and represents as close as possible the underlying population in the region where the nuclear disc dominates. However, the end of the nuclear disc is not trivial to pinpoint, as this structure may gradually fade into the main disc. If the region is too close to the nuclear disc, it may be affected by its young star formation, but if it is too far, it may not map the underlying main disc old star formation. Our methodology used the physical justification based on kinematic maps to select the representative region. Nevertheless, to constrain how much the position of the representative ring could affect our final result, we produce a number of tests with the representative ring at different positions closer to the centre than the position we employed. The results are summarised in Figure~\ref{fig_diffRing_difProf}. 

With the exception of the most central representative ring, for underlying main discs following an exponential light profile, the derived bar ages are constrained to the range $7$--$9$ Gyr.

\subsection{Different main disc profiles -- flat vs. exponential}
\label{apdx_prof}

In order to build the underlying main disc, one has to assume a light profile that describes it. Although galaxy discs are usually assumed as having exponential light profiles, recent studies demonstrate that some cases may follow a flat light profile (e.g., \citealp{zhu2018morphology}; \citealp{breda2020indications}; \citealp{papaderos2021inside}). In order to assess how much such decision may affect our final result, we compare exponential and flat profiles as extreme possible cases. The results are summarised in Figure~\ref{fig_diffRing_difProf}.

For each representative ring position, one sees that the final bar age derived using a flat profile is $0.5$-$1.0$ Gyr older than that derived with an exponential profile (with the exception of the most central representative ring).

Considering the light profile choice together with the position of the representative ring, the derived bar ages vary between $7.0$-$9.5$ Gyr.

\begin{figure*}[ht!]
\centering
\includegraphics[trim={0 0cm 0 0},clip,width=0.9\linewidth]{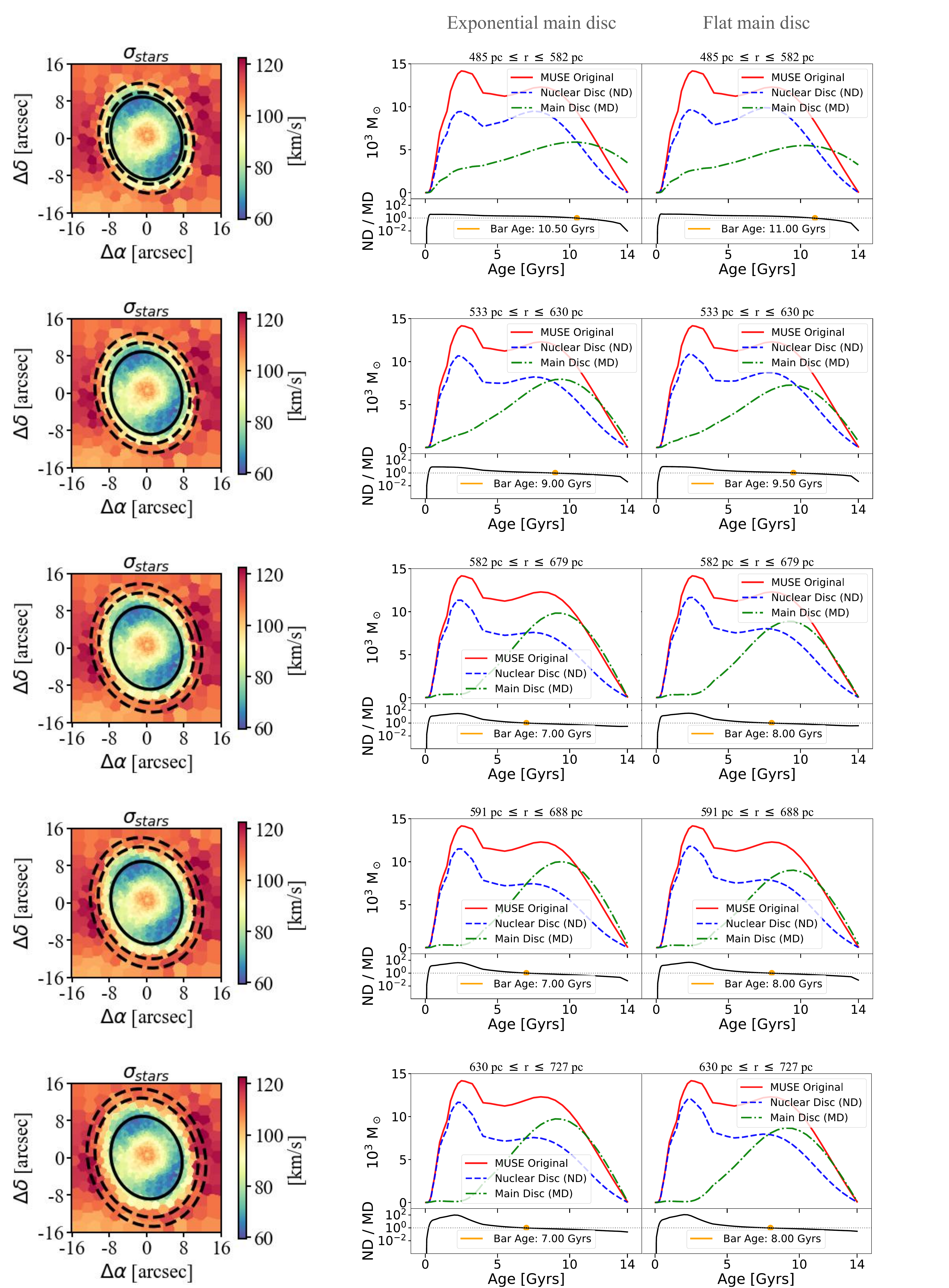}
\end{figure*}

\begin{figure*}[ht!]
\centering
\includegraphics[trim={0 38cm 0 0},clip,width=0.9\linewidth]{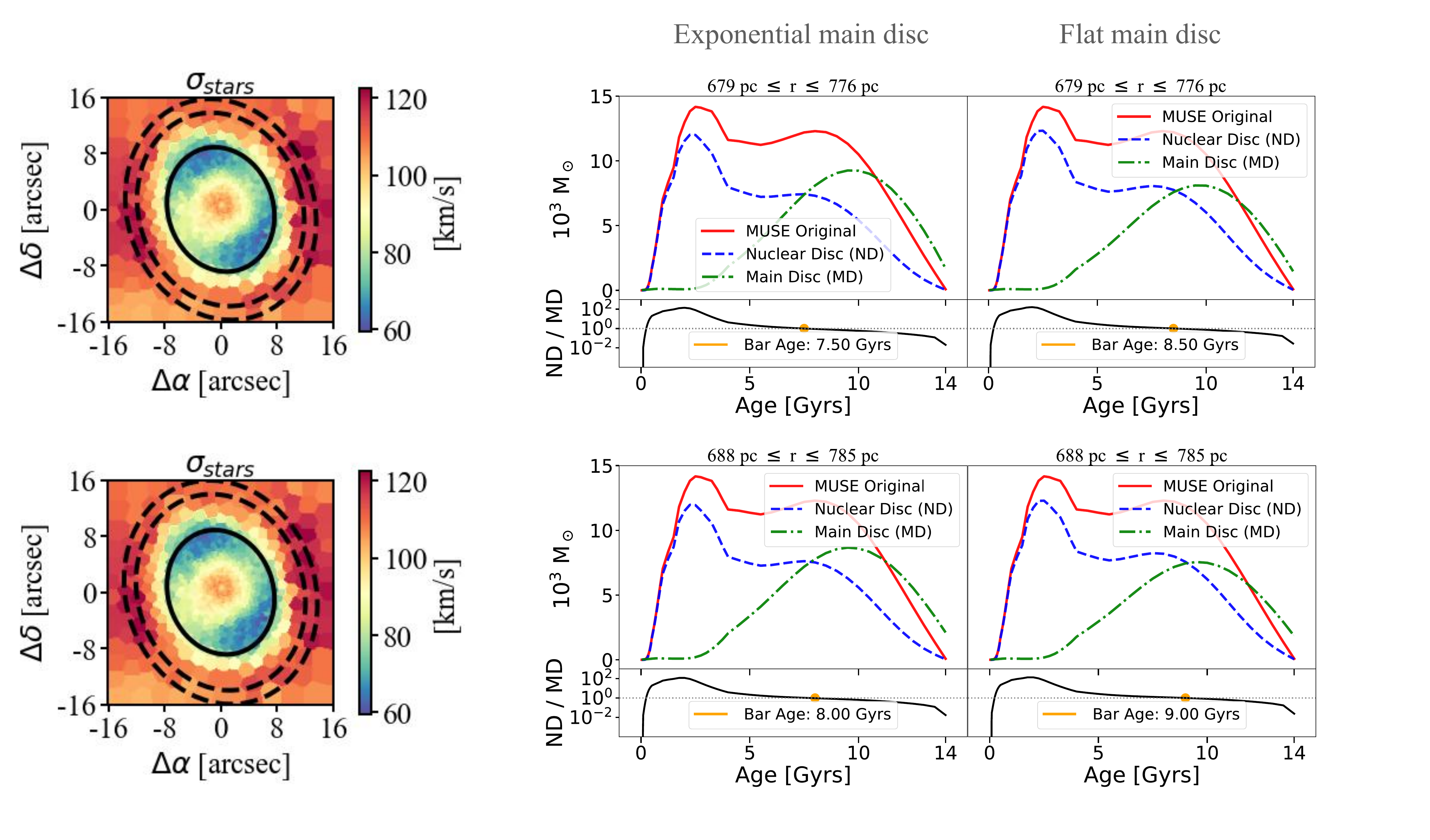}
    \caption{Results for different configurations in the methodology presented in Section~\ref{sec:DisLight}, varying the position of the representative ring (row by row) and the assumed light profile of the main underlying disc. On the first column, we display the stellar velocity dispersion map indicating the representative ring position with the dashed lines (the solid line corresponds to the peak in $v/\sigma$ at the outskirts of the nuclear disc). On the middle and left columns we display results for exponential and flat light profiles, respectively. One can see that the different configurations result in bar ages in the range $7.0$--$9.5$ Gyr, with the exception of the representative ring closest to the centre. The middle panel of the bottom row corresponds to the standard configuration of our methodology.}
    \label{fig_diffRing_difProf}
\end{figure*}

\subsection{Spaxel-by-spaxel analysis}
\label{app_noncollapsed}

\begin{figure}
\centering
\includegraphics[width=1\linewidth]{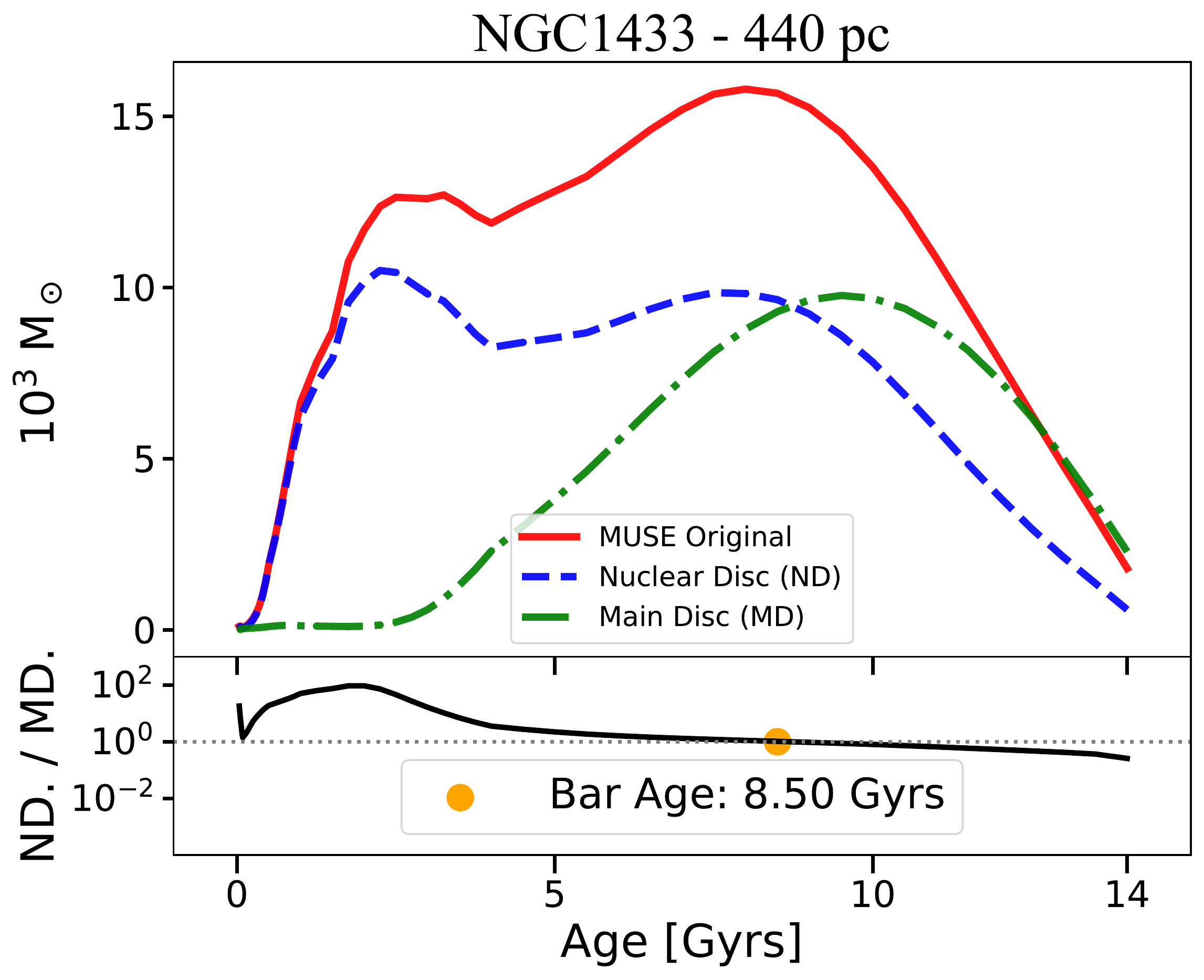}
    \caption{Same as Fig.~\ref{fig_SFH}, but for the test with spaxel-by-spaxel SFHs instead of producing a single collapsed spectrum for each data cube. The SFHs presented are mean SFHs calculated over all spaxels.}
    \label{fig_noncollapsed}
\end{figure}

Considering the elevated signal-to-noise ratio in the TIMER data (\citealp{gadotti2019time}), it would be possible to run the same methodology spaxel-by-spaxel, deriving spatially-resolved SFHs, and then obtaining a mean SFH for each data cube. This would be in contrast to our regular methodology, in which the individual spectra in the data cubes are collapsed into a single spectrum, to then proceed to the derivation of the SFH. The disadvantages of doing so spaxel by spaxel are that the statistical uncertainties for each SFH would be larger and the computational time would increase significantly. Since our goal is to apply the same methodology for the whole TIMER sample, it is too detrimental to require substantial computational time for the analysis of an individual galaxy. Nevertheless, to verify that the results do not differ significantly with the different approaches, we applied the alternative methodology using our data for NGC\,1433. For the spatially-resolved products, we considered the mean SFHs to measure the bar age. The result is shown in Figure~\ref{fig_noncollapsed} and, as one can see, the derived bar age is not substantially different; in fact, the difference is within typical uncertainties in the derivation of stellar ages. With that in mind, we decided to keep our regular methodology employing collapsed spectra to ensure smaller statistical errors and lower computational requirements.

\subsection{pPXF regularisation}
\label{appdx_regul}

\begin{figure*}[ht!]
\centering
\includegraphics[width=1\linewidth]{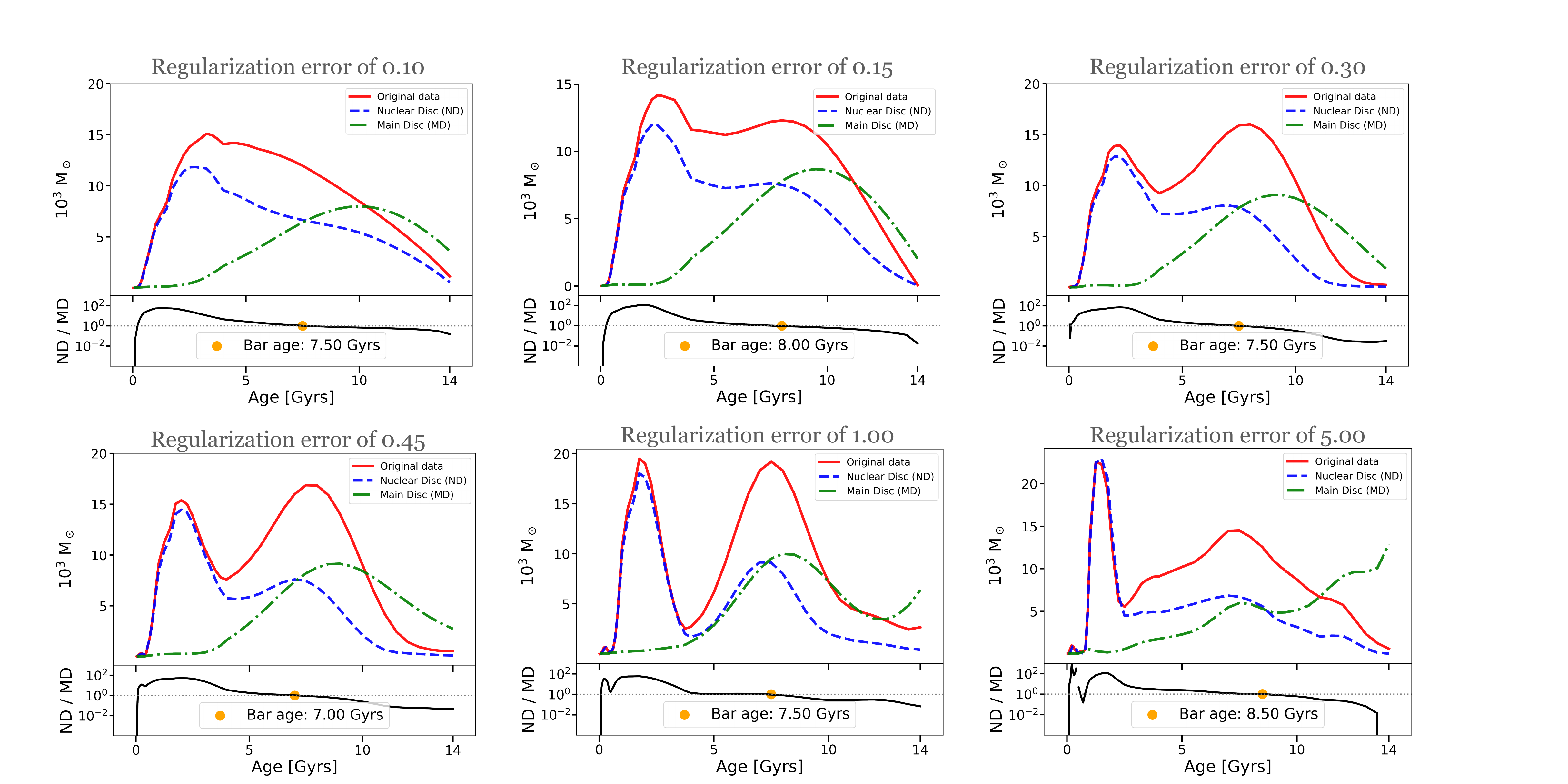}
    \caption{Same as Fig.~\ref{fig_SFH} but for different values of the regularisation error parameter applied by \texttt{pPXF}. The original configuration uses a regularisation error of 0.15. The systematic effect of employing different regularisations is constrained to 0.5 Gyr.}
    \label{fig_difReg}
\end{figure*}

As described in \cite{cappellari2017improving}, regularisation allows one to have the smoothest star formation history solution, without affecting the physical reality of it. In our methodology, we used the \texttt{regul\_err} value of $0.15$ derived by \cite{bittner2020inside} for the TIMER sample following the procedure developed by \cite{mcdermid2015atlas}. Nevertheless, in order to assess how much our final result depends on it, we produce tests with  values of \texttt{regul\_err} in the range $0.1$-$5.0$. The results are summarised in Figure~\ref{fig_difReg}. Since \texttt{regul = 1/regul\_err}, the larger the regularisation error, the less regularised the solution is. 

As one can see, by changing the regularisation error parameter our bar age has variations of $\pm 0.5$ Gyr. Considering the variations due to the representative ring position and the main disc light profile, this leaves us with systematic errors of $^{+1.6}_{-1.1}$ Gyr, after summing in quadrature the maximum variations we find with the four tests on systematic effects.

Finally, the regularization error of $1.00$ leads to a SFH that looks similar to that of the simulation presented in Fig. \ref{fig_sims} (pannel \textbf{e}) which could mislead to an interpretation that the first peak ($\sim 8$ Gyrs) is a contamination of the MD and the second peak ($\sim 5$ Gyrs) is the real bar formation. We point out however that, even though they might look similar at a first glance, one should expect that contamination from the MD would be from the oldest stars, as showed from the different tests for the simulated galaxies (Appendix \ref{test_sims}). Since the oldest peak of the ND is younger than the one from the MD, we do not expect this to be a sign contamination. In addition, this feature is not a well converged feature (as it is not evident for regularisation errors of 0.45 and 5) and is particular to a singular regularization error. As noted in the main body of the paper, the peak at younger ages could arise through a number of mechanisms that induce gas inflow, such as the formation of the boxy/peanut bulge (e.g. \citealt{perez2017observational}) as well as through an interaction or fly-by that removes angular momentum from gas in the disc, causing a burst in gas inflow to the central regions through the bar.

\section{Hydrodynamic Simulations -- Description \& Tests}
\label{sec:simtech}

 The simulation used in Section \ref{subsection3.3} has  both a collisionless (stellar disc and dark matter halo) and collisional (gaseous disc) component and is part of a suite of isolated disc simulations developed to study the evolution of barred galaxies (these will be presented in detail in Fragkoudi \& Bieri, in prep.). Here we describe the main properties of the simulation which are relevant for this study.
The simulation is run using the adaptive mesh refinement (AMR) code RAMSES  \citep{Teyssier2002}. The AMR grid is refined using a quasi-Lagrangian strategy, and has seven refinement levels, with the maximum resolution reached in the simulation being 10\,pc.
The initial conditions of the model are created with the MCMC code DICE \citep{Perretetal2014,Perret2016}.
The total mass of the system is $M_{\rm tot} = 2\times 10^{12} \rm M_{\odot}$, with 98.5\%, 1.425\% and 0.075\% of this distributed in the dark matter, stellar and gaseous components, respectively. The dark matter and stellar components have particles with masses $3.7\times10^4 \rm M_{\odot}$ and $4.3\times10^5 \rm M_{\odot}$, respectively. The dark matter halo has a Navarro-Frenk-White profile \citep{NFW1997} with a scale-length of 3\,kpc, while the stars (gas) are modelled as an exponential disc, with a scale-length of 3\,kpc (4\,kpc) and a scale-height of 150\,pc (50\,pc).

Gas in the simulation cools via atomic and metal-dependant cooling processes. 
%Primordial gas cooling is implemented according to Katz et al. (1996), including collisional excitation, collisional ionisation, recombination and free-free emission, with an additional contribution based on abundances from Suthernald and Dopita (1993). 
%For gas below 10$^4$K, we use the standard prescription implemented in RAMSES which uses the rates from Rosen and Bregman (1995). 
Star formation is modelled as a Schmidt law,
\begin{equation}
    \dot{\rho}_{\star} = \epsilon_{\star}\rho_{\rm gas}/t_{\rm ff},
\end{equation}
where $\dot{\rho}_{\star}$ is the local star formation rate, and $t_{\rm ff} = \sqrt{3\pi/(32G\rho_{\rm gas})}$ is the free-fall time. Star formation is triggered when the gaseous density $\rho_{\rm gas}$ is larger than 1\,cm$^{-3}$ and the temperature is less than 100\,K, with an efficiency of $\epsilon_{\star}=1\%$.
Core-collapse supernova feedback is implemented by assuming that a fraction of the stellar population, $\eta_{\rm SN}=0.2$, will explode as supernovae. The explosion itself is modelled using the mechanical feedback implementation presented in \citet{KimmCen2014} and \citet{Kimmetal2015}. In this implementation, the supernovae explosion is injected into the surrounding interstellar medium according to the phase of the explosion (energy conserving or momentum conserving), i.e. it is injected as momentum or thermal energy, depending on whether the cooling radius is resolved.  More details on the feedback implementation can be found in \citep{KimmCen2014,Kimmetal2015}.

\begin{figure*}
\centering
\includegraphics[width=\linewidth]{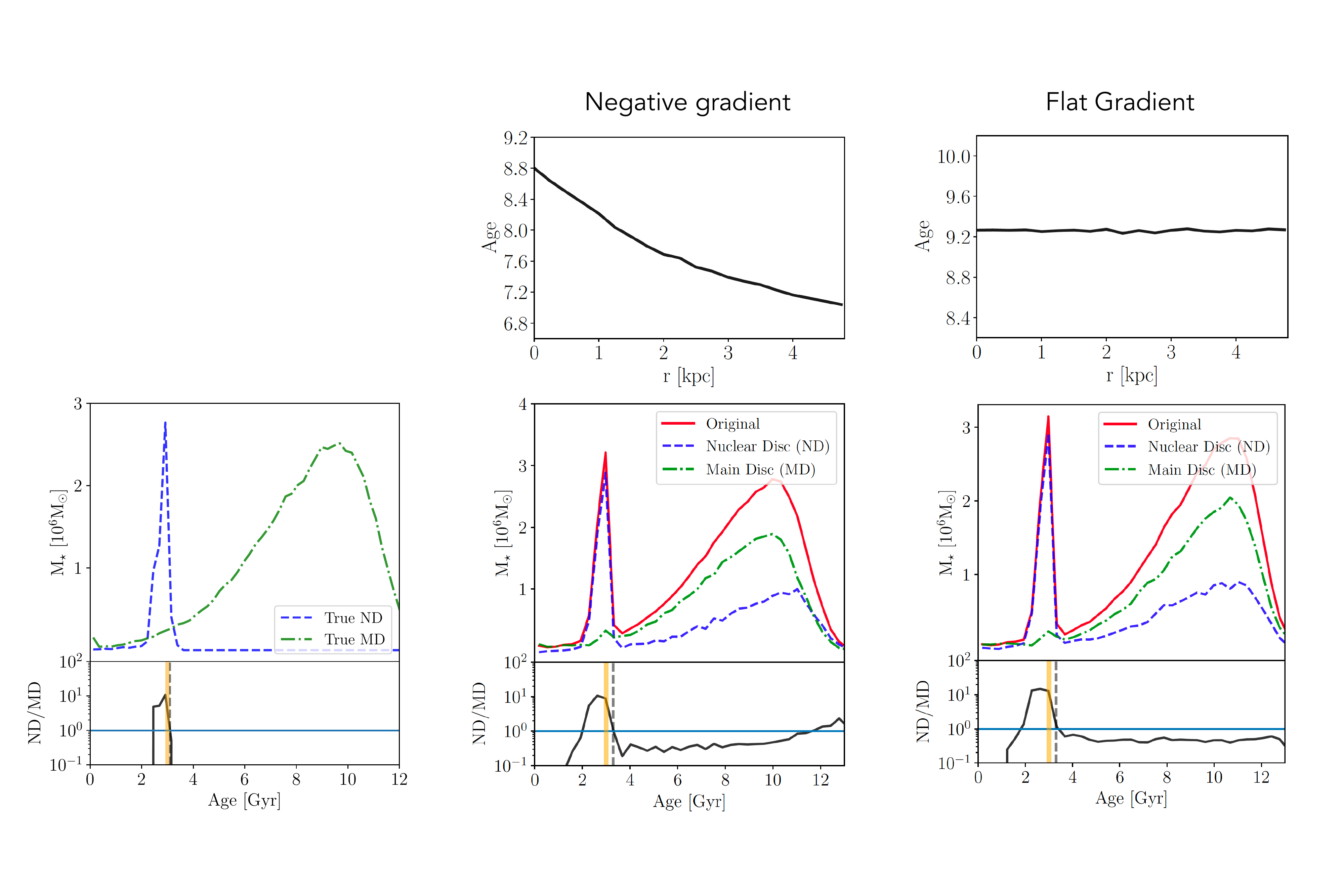}
    \caption{{Testing the bar age criterion and how the age gradient in the underlying main disc affects our methodology for a simulated galaxy:} { In the left panel we show the "true" SFH of the nuclear disc and main disc. The "true" SFH of the nuclear disc is defined as the SFH of all stars formed from the gas pushed to the centre after the bar forms, within the nuclear disc radius $r_{\rm ND}$. The "true" SFH of the main disc is defined as the SFH of all the "old" stars in the  disc, that were present before the bar and nuclear disc formed in the disc, within the same radius, i.e. $r_{\rm ND}$. As one can see, in the bottom panel, the bar age criterion -- the first moment in which ND/MD is above 1 with a positive slope towards younger ages (the vertical grey dashed line) -- is successfully timing the moment the bar is formed (given by the orange vertical line). By comparing this and Fig. \ref{fig_sims} we see that the methodology we present in this paper -- which involves using a representative region around the nuclear disc to model the main disc -- is able to recover the SFH of the nuclear disc, and therefore the bar age.} In the middle and right panels we explore how the age gradient of the old underlying main disc affects our methodology: in the middle panels we show a case where there is a negative age gradient applied to the underlying main disc, while in the right panels we show the effects of a flat age gradient on the results obtained. The top panels show the average age gradient as a function of radius, and the bottom panels show how this affects the SFHs of the original nuclear disc region (red), of the representative SFH of the underlying main disc (green) and of the cleaned nuclear disc (blue), and the lower panels show the ND/MD ratio.}
    \label{fig:agegradienttests}
\end{figure*}

\subsection{Testing the methodology with the hydrodynamic simulations}
\label{test_sims}

In this section we describe some of the tests carried out on the simulations which allow us to assess the effects of some of the assumptions in our methodology. In particular, we test how an assumed underlying age gradient of the old disc, as well as how the location of the ring used to obtain the representative spectrum, affect the results.

{In the left panel of Figure \ref{fig:agegradienttests}, we show what are the ``true'' star formation histories of the nuclear disc and the underlying main disc in the simulations. Then, we show that we can accurately reproduce the time the bar formed using these two values and our chosen criterion for the time of bar formation (i.e., the first moment when the ratio of ND/MD increases above 1 with positive slope towards younger ages; see Sec. \ref{sec:DisLight}). We define the ``true'' SFH of the nuclear disc as the SFH of all stars formed from the gas pushed to the centre after the bar forms within the radius corresponding to the nuclear disc ($r_{\rm ND}$). The ``true'' SFH of the main disc is defined as the SFH of all the "old" stars that were present before the bar and nuclear disc formed in the main disc, within the same radius $r_{\rm ND}$. As we can see, the bar age is well constrained by the moment when ND/MD is above 1 with positive slope towards younger ages for the first time. We can compare ``true'' SFHs to the recovered SFHs from our methodology in Fig. \ref{fig_sims}, where we see that the SFH of the nuclear disc is well recovered, apart from at the oldest ages.} 

%In the left panel of Figure C.1 we show what are the “true” star formation histories of the nuclear disc and of the underlying main disc in the simulations. Then, we show that we can accurately reproduce the time the bar formed using these two values and our chosen criterion for the time of bar formation (i.e., the first moment when the ratio of ND/MD increases above 1 with positive slope towards younger ages, see Section XXX)

\begin{figure*}
\centering
\includegraphics[width=0.9\linewidth]{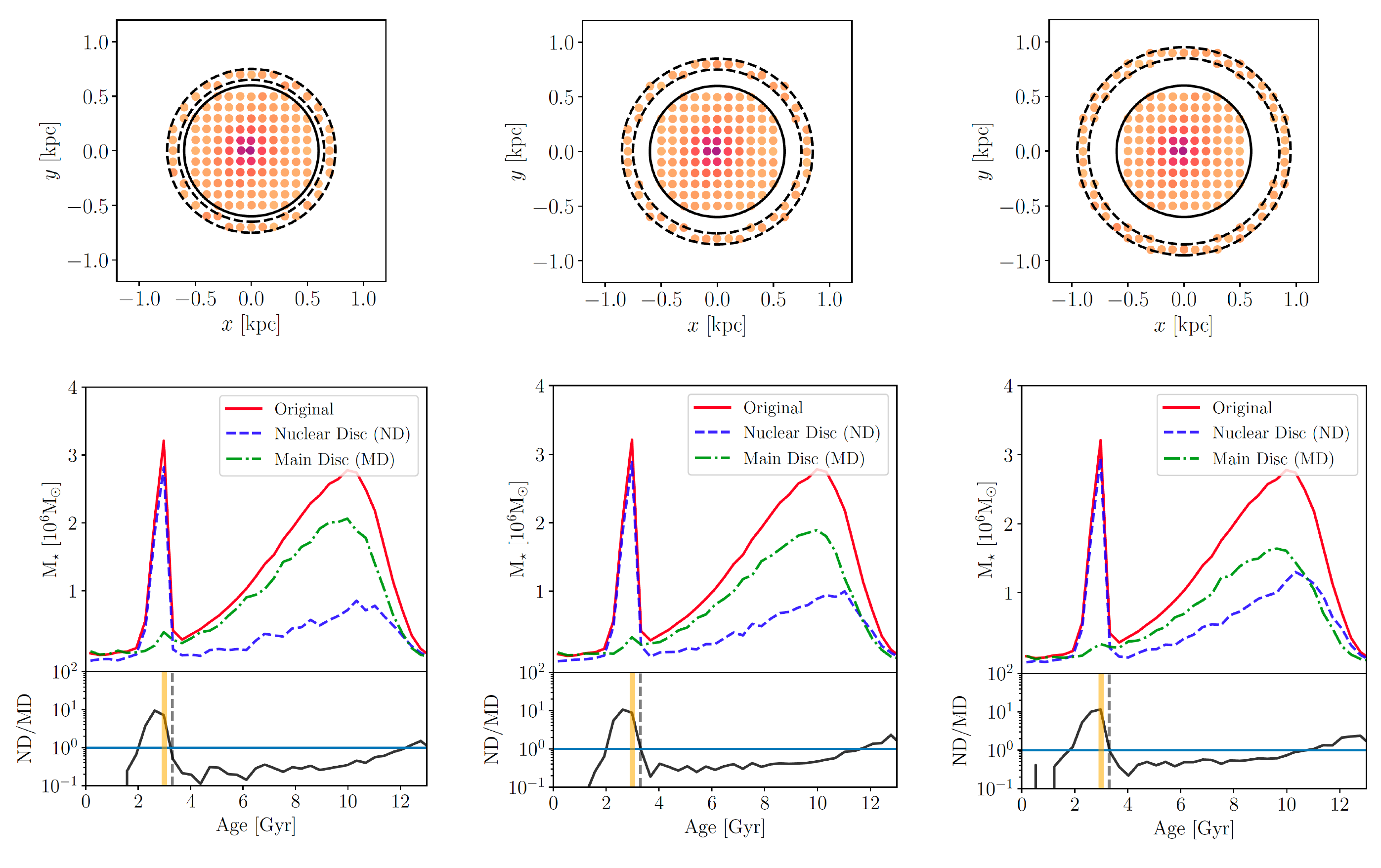}
    \caption{\textbf{Testing the effects of ring location on the methodology for a simulated galaxy:} From left to right we show cases where the ring used to reconstruct the representative SFH of the underlying main disc is placed at larger radii, in a case where the model has a negative age gradient in the underlying main disc. The top rows show the locations of the rings with respect to the nuclear disc region, and the bottom panels show the SFHs. We see that when the ring is placed at larger radii, the representative SFH of the main disc does not fully match the nuclear disc region at the oldest ages. }
    \label{fig:ringloctest}
\end{figure*}

In addition, we test how the underlying age gradient of the main disc in the galaxy can affect the determination of the bar age.  In the right panel of Figure \ref{fig:agegradienttests}, we show a case where there is no age gradient in the underlying main disc of the galaxy. By subtracting the main underlying disc (as obtained from the representative ring around the nuclear disc) from the original data, we obtain a peak in ND/MD that corresponds to the time of bar formation, while before this peak, ND/MD is consistently below $<1$. When a negative age gradient is added to the old disc in the simulation (middle panel), we find similar results, with the difference that, at old ages, ND/MD rises again above one. This happens because there are old stars which are present in the nuclear disc region, which are not present in the ``representative'' main disc region which is at larger radii. Therefore, due to the age gradient, there is a lack of old stars in the ring, which leads to a rise in ND/MD for the oldest populations. This effect will be larger, the larger the age gradient (which will vary from galaxy to galaxy). Due to this, for our criterion of determining the bar age, we explicitly require that ND/MD should be above 1 with a {positive slope towards younger ages} (rather than simply ND/MD $>1$, as this could occur at old ages due to a strong age gradient in the inner galaxy).

In Figure \ref{fig:ringloctest}, we test how the location of the ring affects our methodology, in a case where there is a negative age gradient (if there is no age gradient, the location of the ring does not have a significant effect on the methodology, as long as the SFH of the main disc is appropriately rescaled to take into account the higher surface density in the centre of the galaxy, due to its exponential profile). When the ring is placed at larger radii, the representative SFH will be lacking more of the old stars that are present at smaller radii, which can be seen by the increase in ND/MD $>1$ for old ages, as we move to larger radii (i.e. the middle and right panels of Figure \ref{fig:ringloctest}). Therefore, the representative ring should not be placed at too large radii from the nuclear disc, as the further away it is placed, the more different the stellar populations will be from those of the underlying main disc within the nuclear disc region. As described in Section \ref{sec:DisLight}, we therefore select the radius just outside the nuclear disc, where the average age is oldest (see Figure 3). This ensures we select a region outside the influence of the nuclear disc itself, but without going to larger radii where the stellar population properties of the representative region will be significantly different from those of the underlying main disc in the nuclear disc region.

{We also note that we have applied our methodology to other simulations with the same initial conditions, but with modifications in the star formation and stellar feedback prescriptions, and we were able to recover the bar age reliably also with these changes (i.e., the methodology does not depend on the details of the star formation and stellar feedback prescriptions in the simulation).}

% \begin{figure}
% \centering
% \includegraphics[width=0.8\linewidth]{plots/SFH_Referee_AppC.pdf}
%     \caption{\textbf{The bar age criterium for the "true" SFH of the nuclear disc and the main disc.} The "true" SFH of the nuclear disc is defined as the SFH of all stars formed from the gas pushed to the centre after the bar forms, within the nuclear disc radius $r_{ND}$. The "true" SFH of the main disc is defined as the SFH of all the "old" stars in the  disc, that were present before the bar and nuclear disc formed in the disc, within the same radius, i.e. $r_{ND}$. As one can see, in the bottom panel, the criterium -- the first moment ND/MD is above 1 with positive slope towards younger ages (the vertical grey dashed line) -- is successfully timing the moment the bar is formed (given by the orange vertical line). By comparing this and Fig. \ref{fig_sims} we see that the methodology we present in this paper -- which involves using a representative region around the nuclear disc to model the main disc -- is able to recover the SFH of the nuclear disc, and therefore the bar age.}
%     \label{fig:realSFH}
% \end{figure}

\end{appendix}

%-------------------------------------------------------------------

\end{document}